\documentstyle[prl,aps,preprint,floats,eqsecnum]{revtex}

\begin{document}

\title{Invariant conserved currents in gravity theories with local Lorentz 
and diffeomorphism symmetry}

\author{Yuri N.~Obukhov\footnote{yo@ift.unesp.br}}
\address {Instituto de F\'{\i}sica Te\'orica, Universidade Estadual Paulista,  
Rua Pamplona 145,  01405-900 S\~ao Paulo, Brazil}
\address{Department of Theoretical Physics, Moscow State University, 117234
Moscow, Russia}
\author{Guillermo F.~Rubilar\footnote{grubilar@udec.cl}}
\address {Instituto de F\'{\i}sica Te\'orica, Universidade Estadual Paulista,  
Rua Pamplona 145,  01405-900 S\~ao Paulo, Brazil}
\address{Departamento de F{\'{\i}}sica, Universidad de Concepci\'on,
Casilla 160-C, Concepci\'on, Chile}
\maketitle

\begin{abstract}
We discuss conservation laws for gravity theories invariant under general 
coordinate and local Lorentz transformations. We demonstrate the possibility 
to formulate these conservation laws in many covariant and noncovariant(ly 
looking) ways. An interesting mathematical fact underlies such a diversity: 
there is a certain ambiguity in a definition of the (Lorentz-) covariant 
generalization of the usual Lie derivative. Using this freedom, we develop 
a general approach to construction of {\it invariant} conserved currents
generated by an arbitrary vector field on the spacetime. This is done in any
dimension, for any Lagrangian of the gravitational field and of a (minimally 
or nonminimally) coupled matter field. A development of the ``regularization 
via relocalization" scheme is used to obtain finite conserved quantities for 
asymptotically nonflat solutions. We illustrate how our formalism works 
by some explicit examples.  
\end{abstract}
\bigskip



\section{Introduction}

As it is well known, in the Lagrangian approach every continuous symmetry 
of the action gives rise to a conservation law. This is the substance of 
the Noether theorem which provides a construction of the corresponding 
conserved currents that satisfy certain algebraic and differential 
identities. In particular, the energy and momentum of a physical system
are described by the currents that are generated by the symmetry with 
respect to time and space translations. In a similar way, the angular 
momentum is described by a current related to the spacetime rotations. 

In gravity theory, the definition of energy, momentum and angular momentum
is a nontrivial problem that has a long and rich history (see the reviews 
\cite{Trautman,Faddeev,Szabados}, for example). The difficulty is rooted
deeply in the geometric nature of the gravitational theory and is related
to the equivalence principle which identifies locally gravity and inertia.
As a result, the conservation laws that arise from the general coordinate
(diffeomorphism) invariance normally have the form of covariant, but 
not invariant, equations. The energy-momentum currents in general transform 
nontrivially under a change of the local coordinate system or a local frame. 
In a recent paper \cite{conserved} we have discussed the covariance 
properties of conserved quantities in the framework of the tetrad 
approach to general relativity theory. 

Along with the covariant conservation laws that are formulated in terms 
of the gravitational field variables independently of the additional 
geometric structures on a spacetime manifold, there exist a class of
{\it invariant} conservation laws. The corresponding conserved currents
are usually associated with a vector field that acts on spacetime as
a generator of a certain symmetry of the gravitational configuration. 
However, no systematic and general method for the construction of such 
invariant conserved quantities was ever developed in the literature, at 
least not to our knowledge. For example, quite a long time ago Komar 
\cite{Komar1,Komar2} proposed a very nice formula that gives reasonable
values of the total energy (mass) and angular momentum for asymptotically 
flat configurations. However, although these formulas have proved their 
apparent feasibility, and despite certain achievements 
\cite{Benn,Chrusciel,Katz,Petrov,Fer90,Fer94,Bor94,sard1,sard2,sard3,j1,j2,Wald,De1,De2} 
in explaining their relevance to the concepts of the gravitational energy 
and angular momentum, many important issues remained unclear. In particular,
the freedom in the choice of the gravitational Lagrangian, of the 
gravitational field variables, of the coupling to matter, of the matter 
Lagrangian, of the background geometry (is it needed at all?), of the 
class of relevant vector fields (should they be necessarily Killing vector
fields? why?), all these questions did not have a satisfactory answer 
(in our opinion) in the existing literature. One of the motivations for 
the current study can be formulated as ``understanding Komar". 

Another motivation is related to the interesting work of Aros et al.
\cite{A00a,A00b} who have proposed a new conserved quantity for asymptotically 
anti-de Sitter (AdS) spacetimes. Actually, their result appeared both 
surprising
and confusing to us. The reason is that their conservation law was derived 
from an {\it invariant} Lagrangian with the help of (apparently) {\it
covariant} computations, however, the resulting conserved current (and the 
corresponding total charge) was {\it not a scalar} under local Lorentz 
transformations. How this could happen? Two immediate guesses naturally can 
be formulated. Either the construction \cite{A00a,A00b} is fundamentally 
inconsistent, then it is necessary to find the source of the inconsistency. 
Or, if this result is nevertheless consistent, then perhaps one can improve 
it in such a way that from an invariant Lagrangian, by using covariant 
manipulations, one would be able to derive {\it invariant} conserved 
quantities. At the beginning, we considered the first guess to be most 
probably true. However, a careful analysis has demonstrated that the 
results of \cite{A00a,A00b} are consistent, but can be improved.

In this paper, we present a detailed exposition of the corresponding 
analysis. At the same time, we show that indeed the results of Aros et al.
can be improved along the lines mentioned above. Namely, one of the aims of 
our paper is to systematically investigate the derivation of covariant
and invariant conservation laws for the local Lorentz and the diffeomorphism
symmetry in gravity theories. As a result, we derive new explicitly
{\it invariant} conservation laws for the currents that are true scalars
under general coordinate and local Lorentz transformations. The Lagrangian
approach seems to be most appropriate when one discusses generally covariant
models. We will thus use the Lagrangian framework in this study. Nevertheless,
it is worthwhile to recall that the Hamiltonian approach reveals many other
important aspects, for example the role of the conserved charges in defining
the generators of the corresponding symmetry transformations \cite{Blag}. 
These aspects are discussed in a review \cite{Szabados} (see also the 
references therein). For a covariant Hamiltonian formulation see, in 
particular, \cite{Nester1,Nester2,CN99,Nester3}.

Since both the diffeomorphism and the local Lorentz symmetry are in the
center of our attention, we naturally turn to the so-called Poincar\'e
gauge approach to gravity \cite{Hehl76,Hehl80,Shapiro,Blag,HMMN95,PGrev}
in which both symmetries are naturally realized.
Einstein's general relativity arises as a particular (degenerate) case in
this framework. In the gauge theory of gravity based on the Poincar\'e group 
(the semidirect product of the Lorentz group and the spacetime translations) 
mass (energy-momentum) and spin are treated on an equal footing as the 
sources of the gravitational field. The gravitational gauge potentials 
are the local coframe 1-form $\vartheta^\alpha$ and the 1-form $\Gamma_\alpha
{}^\beta$ of the metric-compatible connection. The absence of nonmetricity
and use of orthonormal frames yield the skew symmetry of the connection,
$\Gamma^{\alpha\beta} = - \Gamma^{\beta\alpha}$. The spacetime manifold 
carries a Riemann-Cartan geometric structure with nontrivial curvature 
and torsion that arise as the corresponding gauge field strengths: $R_\alpha
{}^\beta = d\Gamma_\alpha{}^\beta + \Gamma_\gamma{}^\beta\wedge\Gamma_\alpha
{}^\gamma$ and $T^\alpha = d\vartheta^\alpha + \Gamma_\beta{}^\alpha\wedge
\vartheta^\beta$.

The structure of the paper is as follows. In Sec.~\ref{GLNM} we consider
the consequences of the invariance of a general Lagrangian of the 
gravitational field under local Lorentz transformations of the frames
and under spacetime diffeomorphisms. The Noether identities corresponding
to the Lorentz symmetry are derived. We then demonstrate that these 
identities underlie the possibility to recast the Lie equation for the 
diffeomorphism symmetry in an explicitly covariant form. However, such
a ``covariantization" is essentially non-unique, and it is determined by
what we call a \textit{generalized Lie derivative}. The latter can be 
introduced on a Riemann-Cartan manifold with the help of an arbitrary 
$(1,1)$ tensorial field that has a ``connection"-like transformation law 
under the action of the Lorentz group. These observations are subsequently 
used in Sec.~\ref{diff1} to derive several equivalent forms of the Noether
identities for the diffeomorphism symmetry of the gravitational action.
In Sec.~\ref{matlag} we extend the Lagrange-Noether machinery to the 
Lagrangian of a matter field with an arbitrary coupling to gravity (i.e.,
we allow also for nonminimal coupling).
Sec.~\ref{invcurr} presents the main results of our paper: here we give
the explicit construction of the invariant currents associated with an
arbitrary vector field. The next Sec.~\ref{othercurr} provides a natural
extension of the construction to the case of an arbitrary generalized
Lie derivative. This, in particular, demonstrates the consistency of
the original construction of Aros et al. By applying our general 
construction to the case of the Einstein-Cartan theory in Sec.~\ref{ECT}
we show that the Komar formula arises as a special case of our invariant
current. Sec.~\ref{relocalization} describes the relocalization of the
currents that is induced by the change of the Lagrangian by a boundary
term. In order to demonstrate how the general formalism works, in 
Sec.~\ref{Ex} we apply our derivations to the computation of the conserved 
charges (total mass and angular momentum) for solutions without and
with torsion. In the asymptotically non-flat cases, the conserved charges
turn out to be divergent and we need to regularize them. We show that the 
``regularization via relocalization" method (which we recently used in 
\cite{conserved}) successfully works also here. Finally, 
Sec.~\ref{Discussion} contains a discussion of the results obtained and 
gives an outlook of further possible developments and of open problems. 

Our general notations are as in \cite{HMMN95}. In particular, we use the 
Latin indices $i,j,\dots$ for local holonomic spacetime coordinates and the
Greek indices $\alpha,\beta,\dots$ label (co)frame components. Particular 
frame components are denoted by hats, $\hat 0, \hat 1$, etc. As usual,
the exterior product is denoted by $\wedge$, while the interior product of a 
vector $\xi$ and a $p$-form $\Psi$ is denoted by $\xi\rfloor\Psi$. The vector 
basis dual to the frame 1-forms $\vartheta^\alpha$ is denoted by $e_\alpha$ 
and they satisfy $e_\alpha\rfloor\vartheta^\beta=\delta^\beta_\alpha$. 
Using local coordinates $x^i$, we have $\vartheta^\alpha=h^\alpha_idx^i$ 
and $e_\alpha=h^i_\alpha\partial_i$. We define the volume $n$-form by 
$\eta:=\vartheta^{\hat{0}}\wedge\cdots\wedge\vartheta^{\hat{n}}$. Furthermore,
with the help of the interior product we define $\eta_{\alpha}:=e_\alpha
\rfloor\eta$, $\eta_{\alpha\beta}:=e_\beta\rfloor\eta_\alpha$, $\eta_{\alpha
\beta\gamma}:= e_\gamma\rfloor\eta_{\alpha\beta}$, etc., which are bases for
$(n-1)$-, $(n-2)$- and $(n-3)$-forms, etc., respectively. Finally, 
$\eta_{\alpha_1\cdots \alpha_n} = e_{\alpha_n}\rfloor\eta_{\alpha_1\cdots
\alpha_{n-1}}$ is the Levi-Civita tensor density. The $\eta$-forms satisfy 
the useful identities:
\begin{eqnarray}
\vartheta^\beta\wedge\eta_\alpha &=& \delta_\alpha^\beta\eta ,\\
\vartheta^\beta\wedge\eta_{\mu\nu} &=& \delta^\beta_\nu\eta_{\mu} -
\delta^\beta_\mu\eta_{\nu},\label{veta1}\\ \label{veta}
\vartheta^\beta\wedge\eta_{\alpha\mu\nu}&=&\delta^\beta_\alpha\eta_{\mu\nu
} + \delta^\beta_\mu\eta_{\nu\alpha} + \delta^\beta_\nu\eta_{\alpha\mu},\\
\vartheta^\beta\wedge\eta_{\alpha\gamma\mu\nu}&=&\delta^\beta_\nu\eta_{\alpha
\gamma\mu} - \delta^\beta_\mu\eta_{\alpha\gamma\nu} + \delta^\beta_\gamma
\eta_{\alpha\mu\nu} - \delta^\beta_\alpha\eta_{\gamma\mu\nu},\label{veta2}
\end{eqnarray}
etc. The line element $ds^2 = g_{\alpha\beta}\,\vartheta^\alpha\otimes
\vartheta^\beta$ is defined by the spacetime metric $g_{\alpha\beta}$ 
of signature $(+,-,\cdots,-)$.

\section{General Lagrange-Noether machinery}\label{GLNM}

We work on a $n$-dimensional spacetime manifold. The Poincar\'e gauge
potentials are the coframe $\vartheta^\alpha$ and the Lorentz connection
$\Gamma_\alpha{}^\beta$. We assume that the gravitational Lagrangian 
$n$-form $V$ has the form 
\begin{equation}
V =V(\vartheta^{\alpha},T^{\alpha},R_{\alpha}{}^{\beta})\>,\label{lagrV}
\end{equation}
and that it is \textit{invariant} under local Lorentz transformations. 
As usual, let us introduce, according to the canonical 
prescription, the following translational and rotational {\it gauge 
field momenta} $(n-2)$-forms \cite{foot1}:
\begin{equation} 
H_{\alpha} := -\,{\frac{\partial V}{\partial T^{\alpha}}}\,,\qquad  
H^{\alpha}{}_{\beta} := 
-\,{\frac{\partial V}{\partial R_{\alpha}{}^{\beta}}}\, .\label{HH}
\end{equation}  
Moreover, we define the canonical energy--momentum and spin $(n-1)$-forms 
\begin{equation} 
E_{\alpha} := {\frac{\partial V}{\partial\vartheta^{\alpha}}},\qquad
E^{\alpha\beta}:= 
-\,\vartheta^{[\alpha}\wedge H^{\beta]}, \label{EE}
\end{equation}  
for the gravitational gauge field.

A general variation of the gravitational Lagrangian then reads
\begin{equation}\label{varV}
\delta V =\delta\vartheta^{\alpha}\wedge {\cal E}_\alpha 
+ \delta\Gamma_\alpha{}^{\beta}\wedge{\cal C}^\alpha{}_\beta 
- d\left(\delta\vartheta^{\alpha}\wedge H_\alpha + \delta\Gamma_{\alpha}
{}^{\beta}\wedge H^\alpha{}_\beta\right), \label{var01}
\end{equation}
where we have defined the variational derivatives with respect to (w.r.t.) the
gravitational potentials:
\begin{eqnarray}
{\cal E}_\alpha &:=& {\frac{\delta V}{\delta\vartheta^{\alpha}}}
= -\,DH_\alpha + E_\alpha,\\
{\cal C}^\alpha{}_\beta &:=& {\frac{\delta V}{\delta\Gamma_\alpha
{}^\beta}} = -\,DH^\alpha{}_\beta + E^\alpha{}_\beta.
\end{eqnarray}

For an infinitesimal Lorentz transformation, $\Lambda^\alpha
{}_\beta = \delta^\alpha_\beta + \varepsilon^\alpha{}_\beta$, with 
$\varepsilon_{\alpha\beta} = - \varepsilon_{\beta\alpha}$, we have
\begin{equation}\label{deltaVG}
\delta\vartheta^\alpha = \varepsilon^\alpha{}_\beta\,\vartheta^\beta,
\qquad \delta\Gamma_\beta{}^\alpha = - D\varepsilon^\alpha{}_\beta.
\end{equation}
Substituting this into (\ref{varV}), we find
\begin{equation}
\delta V = \varepsilon^\alpha{}_\beta\left(\vartheta^\beta\wedge 
{\cal E}_\alpha + D{\cal C}^\beta{}_\alpha\right).
\end{equation}
Thus, the local Lorentz invariance of the gravitational Lagrangian, 
$\delta V = 0$, yields the Noether identity
\begin{equation}
D{\cal C}_{\alpha\beta} + \vartheta_{[\alpha}\wedge 
{\cal E}_{\beta]} \equiv 0.\label{Noe1}
\end{equation}

Now, let us derive the consequences of the diffeomorphism invariance
of $V$. Let $f$ be an arbitrary local diffeomorphism on the 
spacetime manifold. It acts with the pull-back map $f^\ast$ on all
the geometrical quantities, and the invariance of the theory means 
that \cite{Kopcz}
\begin{equation}
V(f^\ast\vartheta^\alpha,f^\ast T^\alpha,f^\ast R_\alpha{}^\beta)
= f^\ast(V(\vartheta^{\alpha},T^{\alpha},R_{\alpha}{}^{\beta})).
\end{equation}
Consider an arbitrary vector field $\xi$ and the corresponding
local 1-parameter group of diffeomorphisms $f_t$ generated along 
this vector field. Then, using $f_t$ in the above formula and 
differentiating w.r.t. the parameter $t$, we find the identity
\begin{equation}
(\ell_\xi\vartheta^\alpha)\wedge E_\alpha - (\ell_\xi T^\alpha)\wedge
H_\alpha - (\ell_\xi R_\alpha{}^\beta)\wedge H^\alpha{}_\beta =
\ell_\xi V. \label{lieV1}
\end{equation}
Here the Lie derivative is given on exterior forms by
\begin{equation}
\ell_\xi = d\xi\rfloor + \xi\rfloor d.\label{lie0}
\end{equation}
At first sight, the l.h.s. of (\ref{lieV1}) does not look to be 
invariant under local Lorentz transformations because of the 
usual (not covariant) derivatives in (\ref{lie0}). However, it {\it 
is invariant}. In order to see this, let us recall the property of 
the Lie derivative $\ell_\xi(\varphi\omega) = \varphi\ell_\xi\omega
+ (\xi\rfloor d\varphi)\omega$ which is valid for any exterior
form $\omega$ and any function ($0$-form) $\varphi$. Then, taking 
into account the local Lorentz transformations $\vartheta'^\alpha
= \Lambda^\alpha{}_\beta\vartheta^\beta, E'_\alpha = (\Lambda^{-1}
)^\beta{}_\alpha E_\beta$, $T'^\alpha = \Lambda^\alpha{}_\beta 
T^\beta, H'_\alpha = (\Lambda^{-1})^\beta{}_\alpha H_\beta$,  
$R'_\alpha{}^\beta = \Lambda^\beta{}_\gamma (\Lambda^{-1})^\delta
{}_\alpha R_\delta{}^\gamma$ and $H'^\alpha{}_\beta = \Lambda^\alpha
{}_\gamma (\Lambda^{-1})^\delta{}_\beta H^\gamma{}_\delta$, we 
straightforwardly find that under the action of the local Lorentz 
transformation, the l.h.s. of (\ref{lieV1}) will be shifted by the 
term $(\Lambda^{-1})^\alpha{}_\gamma (\xi\rfloor d\Lambda^\gamma{}_\beta)
\left(D{\cal C}^\beta{}_\alpha + \vartheta^\beta\wedge {\cal E}_\alpha
\right)$. This is zero \textit{in view of the Noether identity} 
(\ref{Noe1}).

Let us now notice that the 1-form $(\Lambda^{-1})^\alpha{}_\gamma 
d\Lambda^\gamma{}_\beta$ is a flat Lorentz connection. Then the above 
observation can be used as follows. We can take an \textit{arbitrary 
Lorentz-valued 0-form} 
$B^{\alpha\beta}(\xi)=-B^{\beta\alpha}(\xi)$ and add a {\it zero term},
\begin{equation}
B_\beta{}^\alpha\left(D{\cal C}^\beta{}_\alpha 
+ \vartheta^\beta\wedge {\cal E}_\alpha\right), \label{zero}
\end{equation}
to the l.h.s. of (\ref{lieV1}). As it is easily verified, this addition is 
{\it equivalent} to the replacement of the usual Lie derivative (\ref{lie0}) 
with a {\it generalized Lie derivative} $L_\xi:=\ell_\xi + B_\beta{}^\alpha
\rho^\beta{}_\alpha$ when applied to coframe, torsion and curvature. This 
generalized derivative will be \textit{covariant} provided $B_\alpha{}^\beta$ 
transforms according to (\ref{TB}). Here $\rho^\beta{}_\alpha$ denote the 
corresponding Lorentz generators for the object on which the derivative acts. 

There are many options for the choice of the generalized covariant Lie 
derivative. One family of generalized Lie derivatives corresponds to the 
case in which $B_\beta{}^\alpha=\xi\rfloor A_\beta{}^\alpha$, where $A_\beta
{}^\alpha$ is an {\it arbitrary Lorentz connection}. For this family, one can 
verify that $\ell_\xi + \xi\rfloor A_\beta{}^\alpha\rho^\beta{}_\alpha=\xi
\rfloor\stackrel{A}{D}+\stackrel{A}{D}\xi\rfloor$, where $\stackrel{A}{D}
:=d+A_\beta{}^\alpha\rho^\beta{}_\alpha$ denotes the covariant derivative 
defined by the connection $A$. We will consider the following two
possibilities: 
\begin{eqnarray}
{\hbox{\L}}_\xi &:=&\ell_\xi + \xi\rfloor\Gamma_\beta{}^\alpha
\rho^\beta{}_\alpha=\xi\rfloor D + D\xi\rfloor,\label{lie1}\\
{\stackrel{\{\,\}}{\hbox{\L}}}_\xi &:=&\ell_\xi + \xi\rfloor{\stackrel{\{\,\}}
{\Gamma}}_\beta{}^\alpha\rho^\beta{}_\alpha = \xi\rfloor{\stackrel{\{\,\}}{D}}
+ {\stackrel{\{\,\}}{D}}\xi\rfloor.\label{lie2}
\end{eqnarray}
Here $\Gamma_\alpha{}^\beta$ is the dynamical Lorentz connection of the 
theory, whereas ${\stackrel{\{\,\}}{\Gamma}}_\alpha{}^\beta$ is the 
Riemannian connection (i.e., the anholonomic form of the Christoffel 
symbols). We refer the reader to Appendix~\ref{app1} for further details.

Yet another possibility satisfying the condition (\ref{TB}), that
is not directly related to a connection, is given by
\begin{equation}
{\cal L}_\xi := \ell_\xi - \Theta_\beta{}^\alpha\rho^\beta{}_\alpha,
\label{lie3}
\end{equation} 
i.e., $B_\alpha{}^\beta:=-\Theta_\alpha{}^\beta$, where
\begin{equation}\label{theta}
\Theta^{\alpha\beta} := e^{[\alpha}\rfloor\ell_\xi\vartheta^{\beta]}. 
\end{equation}

This our third choice of generalized covariant Lie derivative is 
very different from the other two, as its existence is specific  
for the models with a coframe (tetrad) field (and a metric). 
We will call (\ref{lie3}) a {\it Yano derivative} since its counterpart 
for the linear group was first introduced in \cite{Yano}. See also 
\cite{Mielke01} for similar, but different definitions.

\section{Gravitational Noether identities for diffeomorphism symmetry}
\label{diff}

All the above three choices of the covariant Lie derivative are useful.
The most common is the first option (\ref{lie1}) which directly yields 
covariant Noether identities \cite{HMMN95}. (For a general discussion
of the Noether theorems in models with local symmetries see \cite{Bar}).

\subsection{Covariant Noether identities}\label{diff1}

Using the covariant Lie derivative ${\hbox{\L}}_\xi$  we recast 
(\ref{lieV1}) into
\begin{equation}
({\hbox{\L}}_\xi\vartheta^\alpha)\wedge E_\alpha - ({\hbox{\L}}_\xi
T^\alpha)\wedge H_\alpha - ({\hbox{\L}}_\xi R_\alpha{}^\beta)\wedge 
H^\alpha{}_\beta - {\hbox{\L}}_\xi V = A + dB =0,\label{lieV2}
\end{equation}
from where (with $\xi^\alpha := \xi\rfloor\vartheta^\alpha$ denoting 
the components of the vector field $\xi$), we find
\begin{eqnarray}
A &=& \xi^\alpha\left(-\,D{\cal E}_\alpha + e_\alpha\rfloor T^\beta
\wedge{\cal E}_\beta + e_\alpha\rfloor R_\gamma{}^\beta\wedge
{\cal C}^\gamma{}_\beta\right),\label{A1}\\
B &=& \xi^\alpha\left(E_\alpha - e_\alpha\rfloor T^\beta
\wedge H_\beta - e_\alpha\rfloor R_\gamma{}^\beta\wedge
H^\gamma{}_\beta - e_\alpha\rfloor V\right).\label{B} 
\end{eqnarray}
Since the diffeomorphism invariance holds for {\it arbitrary}
vector fields $\xi$, $A$ and $B$ must necessarily vanish and thus 
we find the familiar Noether identities:
\begin{eqnarray}
D{\cal E}_\alpha &\equiv& e_\alpha\rfloor T^\beta
\wedge{\cal E}_\beta + e_\alpha\rfloor R_\gamma{}^\beta\wedge
{\cal C}^\gamma{}_\beta,\label{NoeD1}\\
E_\alpha &\equiv& e_\alpha\rfloor V + e_\alpha\rfloor T^\beta
\wedge H_\beta + e_\alpha\rfloor R_\gamma{}^\beta\wedge
H^\gamma{}_\beta. \label{n02}
\end{eqnarray}

\subsection{Covariant Noether identities: another face}\label{diff2}

The second choice (\ref{lie2}) brings (\ref{lieV1}) into a
similar form:
\begin{equation}\label{lieV3}
({\stackrel{\{\,\}}{\hbox{\L}}}_\xi\vartheta^\alpha)\wedge E_\alpha - 
({\stackrel{\{\,\}}{\hbox{\L}}}_\xi T^\alpha)\wedge H_\alpha - 
({\stackrel{\{\,\}}{\hbox{\L}}}_\xi R_\alpha{}^\beta)\wedge H^\alpha
{}_\beta - {\stackrel{\{\,\}}{\hbox{\L}}}_\xi V = A' + dB =0.
\end{equation}
For  $B$ we still find (\ref{B}), whereas $A'$ is different,
acquiring an additional piece proportional to the rotational
Noether identity, namely
\begin{equation}
A' = \xi^\alpha\left[-\,D{\cal E}_\alpha + e_\alpha
\rfloor T^\beta\wedge{\cal E}_\beta + e_\alpha\rfloor R_\gamma
{}^\beta\wedge {\cal C}^\gamma{}_\beta + (e_\alpha\rfloor 
K_\gamma{}^\beta)\left(\vartheta^\gamma\wedge{\cal E}_\beta
+ D{\cal C}^\gamma{}_\beta\right)\right]. \label{A2}
\end{equation}
In accordance with the above general analysis, the difference
of (\ref{lieV3}) and (\ref{lieV2}) is proportional to the 
difference of the connections
\begin{equation}
K_\alpha{}^\beta := {\stackrel{\{\,\}}{\Gamma}}_\alpha{}^\beta
- \Gamma_\alpha{}^\beta.\label{K}
\end{equation}
This quantity is known as contortion 1-form. In particular, 
the torsion is recovered from it as $T^\alpha = K^\alpha{}_\beta
\wedge\vartheta^\beta$. The corresponding curvature 2-forms are
related via 
\begin{equation}\label{RR}
R_\alpha{}^\beta = {\stackrel{\{\,\}}{R}}_\alpha{}^\beta - {\stackrel
{\{\,\}}{D}}K_\alpha{}^\beta + K_\gamma{}^\beta\wedge K_\alpha{}^\gamma.
\end{equation}
Combining together the terms with ${\cal E}_\alpha$, with the help
of (\ref{K}) and (\ref{RR}) we can recast (\ref{A2}) into
\begin{equation}
A' = \xi^\alpha\left[-\,{\stackrel{\{\,\}}{D}}\left(
{\cal E}_\alpha - {\cal C}^\gamma{}_\beta e_\alpha\rfloor 
K_\gamma{}^\beta\right) + {\cal C}^\gamma{}_\beta\wedge
\left(e_\alpha\rfloor{\stackrel{\{\,\}}{R}}_\gamma{}^\beta 
- {\stackrel{\{\,\}}{\hbox{\L}}}_\alpha K_\gamma{}^\beta\right)
\right]. \label{A3}
\end{equation}
Again, since the diffeomorphism invariance holds for an
arbitrary vector field $\xi$, we recover the corresponding
Noether identity in the alternative form
\begin{equation}
{\stackrel{\{\,\}}{D}}\left({\cal E}_\alpha - {\cal C}^\gamma{}_\beta
e_\alpha\rfloor K_\gamma{}^\beta\right) \equiv \left(e_\alpha\rfloor
{\stackrel{\{\,\}}{R}}_\gamma{}^\beta - {\stackrel{\{\,\}}
{\hbox{\L}}}_\alpha K_\gamma{}^\beta\right)\wedge 
{\cal C}^\gamma{}_\beta.\label{NoeD2}
\end{equation}
The identities (\ref{NoeD1}) and (\ref{NoeD2}) are 
equivalent, but in contrast to the usual (\ref{NoeD1}), the 
alternative form (\ref{NoeD2}) is less known. It was derived 
previously in \cite{PGrev} using a different method. 

\subsection{``Noncovariant" Noether identities}\label{diff0}

Besides the above choices, it is still possible to work with
the noncovariant ordinary Lie derivative (\ref{lie0}). Then we start 
directly with the identity (\ref{lieV1}) which we recast into 
\begin{equation}
(\ell_\xi\vartheta^\alpha)\wedge E_\alpha - (\ell_\xi T^\alpha)
\wedge H_\alpha - (\ell_\xi R_\alpha{}^\beta)\wedge H^\alpha{}_\beta
- \ell_\xi V = {\cal A} + dB = 0. \label{lieV4}
\end{equation}
Again $B$ is given by (\ref{B}), but $A$ changes into
\begin{eqnarray}
{\cal A} &=& \xi^\alpha\left[-\,D{\cal E}_\alpha + e_\alpha
\rfloor T^\beta\wedge{\cal E}_\beta + e_\alpha\rfloor R_\gamma
{}^\beta\wedge {\cal C}^\gamma{}_\beta - (e_\alpha\rfloor 
\Gamma_\gamma{}^\beta)\left(\vartheta^\gamma\wedge{\cal E}_\beta
+ D{\cal C}^\gamma{}_\beta\right)\right]\nonumber\\
&=& \xi^\alpha\left[-\,d{\cal E}_\alpha + (e_\alpha\rfloor d
\vartheta^\beta)\wedge{\cal E}_\beta + (e_\alpha\rfloor d\Gamma
{}_\gamma{}^\beta)\wedge {\cal C}^\gamma{}_\beta - (e_\alpha\rfloor
\Gamma_\gamma{}^\beta)\,d{\cal C}^\gamma{}_\beta\right].\label{A4}
\end{eqnarray}
As a result of the diffeomorphism invariance for an arbitrary $\xi$
we find yet another form of the Noether identity (\ref{NoeD1}), namely
\begin{equation}
d\left({\cal E}_\alpha + {\cal C}^\gamma{}_\beta\,e_\alpha\rfloor
\Gamma_\gamma{}^\beta\right) \equiv (\ell_{e_\alpha}\vartheta^\beta)
\wedge{\cal E}_\beta + (\ell_{e_\alpha}\Gamma_\gamma{}^\beta)\wedge 
{\cal C}^\gamma{}_\beta.\label{NoeD3}
\end{equation}
One may call this form of the Noether identity ``noncovariant"
since it explicitly involves the noncovariant gravitational field
potentials $\Gamma_\alpha{}^\beta$ (and not the corresponding field 
strengths) and the ordinary (noncovariant) Lie derivative. However, 
just like the noncovariantly looking expression (\ref{lieV1}), the 
identity (\ref{NoeD3}) is in fact {\it covariant}. Moreover, all 
the three forms (\ref{NoeD1}), (\ref{NoeD2}) and (\ref{NoeD3}) are 
completely equivalent. 

\section{Matter: dynamics and Noether identities}\label{matlag}

Matter fields can be represented by scalar-, tensor- or spinor-valued 
forms of some rank, and we will denote all of them collectively as $\Psi^A$, 
where the superscript $\scriptstyle A$ indicates the appropriate index 
(tensor and/or spinor) structure. We will assume that the matter fields 
$\Psi^A$ belong to the space of some (reducible, in general) representation
of the Lorentz group. For an infinitesimal Lorentz transformation 
(\ref{deltaVG}) the matter fields transform as
\begin{equation}\label{deltaP}
\Psi^{\prime A}= \Psi^A + \delta\Psi^A,\quad\quad\delta\Psi^A=
\varepsilon^\beta{}_\alpha\,(\rho^\alpha{}_\beta)^A{}_B\Psi^B.
\end{equation}
Here $(\rho^\alpha{}_\beta)^A{}_B$ denote the corresponding matrices of 
generators of the Lorentz group. For definiteness, we consider the case 
when the matter field $\Psi^A$ is a $0$-form on the spacetime manifold. 
This includes scalar and spinor fields of any rank \cite{gaugematter}.

\subsection{Lagrangian and field equations}

We assume that the matter Lagrangian $n$-form $L$ depends most generally 
on $\Psi^A$, $d\Psi^A$ and the gravitational potentials $\vartheta^{\alpha}$,
$\Gamma_{\alpha}{}^{\beta}$. According to the {\it minimal coupling} 
prescription, derivatives of the gravitational potentials are not permitted. 
We usually adhere to this principle. However, Pauli--type terms 
and Jordan--Brans--Dicke--type terms may occur in phenomenological 
models or in the context of a symmetry breaking mechanism. Also the Gordon 
decomposition of the matter currents and the discussion of the gravitational 
moments necessarily requires the inclusion of Pauli--type terms, see 
\cite{Hehl4,Fermion}. Therefore, we develop our Lagrangian formalism in 
sufficient generality in order to cope with such models by including in the 
Lagrangian also the derivatives $d\vartheta^{\alpha}$, and $d\Gamma_{\alpha}
{}^{\beta}$ of the gravitational potentials. In view of the assumed invariance 
of $L$ under local Lorentz transformations, the derivatives can enter only 
in a covariant form, namely
\begin{equation}\label{Lmat}
L=L(\Psi^A, D\Psi^A, \vartheta^\alpha, T^\alpha, R_\alpha{}^\beta)\,. 
\end{equation} 
That is, the derivatives of the matter fields appear only in a covariant
combination, $D\Psi^A = d\Psi^A + \Gamma_\alpha{}^\beta\wedge(\rho^\alpha
{}_\beta)^A{}_B\,\Psi^B$, whereas the derivatives of the coframe and 
connection can only appear via the torsion and the curvature 2-forms.

For the total variation of the matter Lagrangian with respect to the 
material and the gravitational fields, we find
\begin{eqnarray}
\delta L &=& \delta\vartheta^\alpha\wedge{\frac {\partial L}{\partial
\vartheta^\alpha}} + \delta T^\alpha\wedge{\frac {\partial L}{\partial 
T^\alpha}} + \delta R_{\alpha}{}^{\beta}\wedge{\frac {\partial L} 
{\partial R_{\alpha}{}^{\beta}}}\nonumber\\ 
&&\qquad +\,\delta\Psi^A\wedge{\frac {\partial L} {\partial\Psi^A}}+\delta
(D\Psi^A)\wedge{\frac {\partial L} {\partial D\Psi^A}}\label{vard0}\\
&=& \delta\vartheta^{\alpha}\wedge\Sigma_\alpha +
\delta\Gamma_\alpha{}^{\beta}\wedge\tau^\alpha{}_\beta 
+\delta\Psi^A\wedge{\frac{\delta L}{\delta\Psi^A}}\nonumber\\
&&\quad + d\left[\delta\vartheta^{\alpha}\wedge{\frac
{\partial L}{\partial T^{\alpha}}}+ \delta\Gamma_{\alpha}{}^{\beta}
\wedge{\frac{\partial L}{\partial R_{\alpha}{}^{\beta}}}+ \delta\Psi^A
\wedge{\frac{\partial L}{\partial D\Psi^A}}\right]\,.\label{vard1}
\end{eqnarray}
Here, as usual, we denote the covariant {\it variational derivative} 
of $L$ with respect to the matter field $\Psi^A$ as
\begin{equation}
{\frac {\delta L} {\delta\Psi^A}} :={\frac {\partial  L}{\partial\Psi^A}} 
- D\,{\frac {\partial L}{\partial (D\Psi^A)}}, \label{psi0}
\end{equation}
and the {\it canonical currents} of energy-momentum and spin are defined,
respectively, by \cite{F2}
\begin{eqnarray}
\Sigma_{\alpha}&:=& {\frac {\delta L}{\delta\vartheta^{\alpha}}} =  
{\frac {\partial L}{\partial\vartheta^{\alpha}}} + D\,{\frac {\partial L}
{\partial T^{\alpha}}}\, ,\label{sigC0}\\
\tau^\alpha{}_\beta &:=& {\frac {\delta L}{\delta\Gamma_\alpha{}^\beta}} 
= (\rho^\alpha{}_\beta)^A{}_B\,\Psi^B\wedge{\frac {\partial L} 
{\partial (D\Psi^A)}} + \vartheta^{[\alpha}\wedge 
{\frac {\partial L}{\partial T^{\beta]}}} + D\,{\frac {\partial L}
{\partial R_\alpha{}^\beta}}\,.\label{spin0}
\end{eqnarray}

The principle of stationary action $\delta\int V^{\rm tot} =0$, 
$V^{\rm tot}:=V + L$, for the coupled system of the gravitational and 
material fields, see (\ref{varV}) and (\ref{vard1}), then yields the 
dynamical equations:
\begin{eqnarray}
{\frac {\delta L} {\delta\Psi^A}} &=& 0,\label{matter}\\
{\cal E}_\alpha + \Sigma_\alpha &=& 0,\label{einstein}\\
{\cal C}^\alpha{}_\beta + \tau^\alpha{}_\beta &=& 0.\label{cartan}
\end{eqnarray}

\subsection{Lorentz symmetry}

It is straightforward to derive the Noether identities following from the 
invariance of $L$ under the local Lorentz group and under diffeomorphisms. 
Using the infinitesimal transformations (\ref{deltaVG}) and (\ref{deltaP}), 
we find for the {\it Lorentz symmetry}:
\begin{eqnarray}
\delta L &=&\varepsilon^\beta{}_\alpha\,\Bigl(\vartheta^{\alpha}
\wedge\Sigma_{\beta} + D\tau^\alpha{}_\beta + (\rho^\alpha{}_\beta
)^A{}_B\Psi^B\wedge {\frac{\delta L}{\delta\Psi^A}}\Bigr)\nonumber\\
&& + \,d\left[\varepsilon^\beta{}_\alpha\left(-\,\tau^\alpha{}_\beta
+ (\rho^\alpha{}_\beta)^A{}_B\Psi^B\wedge{\frac {\partial L}{\partial 
D\Psi^A}}+ \vartheta^\alpha\wedge{\frac{\partial L}{\partial T^\beta}} 
+D\,{\frac{\partial L}{\partial R_\alpha{}^\beta}}\right)\right] =0.
\label{invL}
\end{eqnarray}
The term in the total derivative vanishes identically in view of the 
definition (\ref{spin0}) of the spin current $\tau^\alpha{}_\beta$. Then, 
from the arbitrariness of $\varepsilon^\alpha{}_\beta$, we find the 
corresponding Noether identity:
\begin{equation}
D\tau_{\alpha\beta} +  \vartheta_{[\alpha}\wedge\Sigma_{\beta]}\equiv 
-\,(\rho_{\alpha\beta})^A{}_B\Psi^B\wedge{\frac{\delta L}{\delta\Psi^A}}
\cong 0\,.\label{Noe2}
\end{equation}
With $\cong$ we denote ``weak identities'', i.e., an identity valid assuming 
that the \textit{matter} field equations are satisfied.

\subsection{Diffeomorphism symmetry}

For a {\it diffeomorphism} generated by an arbitrary vector field $\xi$,
we find directly from (\ref{vard0}):
\begin{eqnarray}
\ell_\xi L &=& (\ell_{\xi}\vartheta^\alpha)\wedge{\frac{\partial L} 
{\partial\vartheta^\alpha}}+ (\ell_{\xi} T^\alpha)\wedge{\frac{\partial L} 
{\partial T^\alpha}} + (\ell_{\xi} R_{\alpha}{}^{\beta})\wedge{\frac
{\partial L} {\partial R_{\alpha}{}^{\beta}}}\nonumber\\ 
&&\quad +(\ell_\xi\Psi^A)\wedge{\frac{\partial L} {\partial\Psi^A}}
+ (\ell_{\xi}D\Psi^A)\wedge {\frac{\partial L} {\partial D\Psi^A}}\,.
\label{vard2}
\end{eqnarray}
In complete analogy to the purely gravitational case, we have a wide 
choice of covariant Lie derivatives which we can use in the above
identity instead of the ordinary Lie derivative $\ell_\xi$. However, we
will not repeat once again the detailed analysis of all possibilities, 
taking into account the equivalence of the final results. Instead, here
we give details for the covariant option (\ref{lie1}) only. We then find
\begin{eqnarray}
-\,{\hbox{\L}}_\xi L &+& ({\hbox{\L}}_{\xi}\vartheta^\alpha)\wedge
{\frac{\partial L} {\partial\vartheta^\alpha}} + ({\hbox{\L}}_{\xi} 
T^\alpha)\wedge{\frac{\partial L} {\partial T^\alpha}} + ({\hbox{\L}
}_{\xi} R_{\alpha}{}^{\beta})\wedge{\frac{\partial L} {\partial R_{\alpha}
{}^{\beta}}}\nonumber\\ &&\quad +({\hbox{\L}}_\xi\Psi^A)\wedge{\frac
{\partial L} {\partial\Psi^A}}+ ({\hbox{\L}}_{\xi}D\Psi^A)\wedge 
{\frac{\partial L} {\partial D\Psi^A}} = A + dB = 0,\label{vard3}
\end{eqnarray}
where, after some algebra, we find
\begin{eqnarray}
A &=& -\,(\xi\rfloor\vartheta^{\alpha})D{\frac{\delta L}{\delta
\vartheta^{\alpha}}} + (\xi\rfloor T^{\alpha})\wedge{\frac{\delta L}
{\delta\vartheta^{\alpha}}} + (\xi\rfloor R_\beta{}^{\gamma})\wedge
{\frac{\delta L}{\delta \Gamma_{\beta}{}^{\gamma}}}\nonumber\\
&& +\,(\xi\rfloor D\Psi^A)\,\wedge{\frac{\delta L}{{\delta\Psi^A}}},\\
B &=& (\xi\rfloor\vartheta^\alpha)\,{\frac{\partial L}{\partial
\vartheta^{\alpha}}} + (\xi\rfloor T^\alpha)\wedge{\frac{\partial L}
{\partial T^\alpha}} + (\xi\rfloor R_{\alpha}{}^{\beta})\wedge
{\frac{\partial L}{\partial R_{\alpha}{}^{\beta}}}\nonumber\\
&& +\,(\xi\rfloor D\Psi^A)\wedge{\frac{\partial L}{\partial D\Psi^A}}
- \xi\rfloor L.
\end{eqnarray}
Using the fact that $\xi$ is a pointwise arbitrary vector field, we derive
the following Noether identities from $B = 0$ and $A =0$, respectively:
\begin{eqnarray}
\Sigma_\alpha &\equiv& e_\alpha\rfloor L - (e_\alpha\rfloor D\Psi^A)
\wedge {\frac{\partial L}{\partial D\Psi^A}}\nonumber\\
&&+ D{\frac{\partial L}{\partial T^\alpha}}-
(e_{\alpha}\rfloor T^\beta)\wedge{\frac{\partial L}{\partial T^\beta}} 
- (e_{\alpha}\rfloor R_{\beta}{}^{\gamma})\wedge 
{\frac{\partial L}{\partial R_{\beta}{}^{\gamma}}},\label{momC}
\end{eqnarray}
and
\begin{eqnarray}
D\Sigma_\alpha &\equiv & (e_\alpha\rfloor T^\beta)\wedge\Sigma_\beta
+ (e_\alpha\rfloor R_{\beta}{}^{\gamma})\wedge\tau^{\beta}{}_{\gamma}
+ \,w_{\alpha}\nonumber\\ 
& \cong & (e_\alpha\rfloor T^\beta)\wedge\Sigma_\beta + (e_\alpha
\rfloor R_\beta{}^\gamma)\wedge\tau^\beta{}_\gamma,\label{conmomC}
\end{eqnarray}
where 
\begin{equation}\label{wa}
w_{\alpha}:=(e_\alpha\rfloor D\Psi^A)\,{\frac{\delta L}{\delta\Psi^A}}\cong 0.
\end{equation}

Equation (\ref{momC}) yields the explicit form of the {\it canonical
energy-momentum of matter}, with the first line representing the result 
known in the context of the {\it special}--relativistic classical field 
theory. The second line in (\ref{momC}) accounts for the possible Pauli 
terms as well as for Lagrange multiplier terms in the variations 
with constraints and it is absent for the case of minimal coupling.
The first line in (\ref{conmomC}) is given in the {\it strong} form, 
without using the field equations for matter (\ref{matter}).

In complete analogy to Sec.~\ref{diff2} and Sec.~\ref{diff0}, it is 
possible to derive the ``Riemannian" and the ``non-covariant" versions 
of the Noether identity (\ref{conmomC}). By using instead of (\ref{lie1}) 
the covariant Lie derivative (\ref{lie2}) we then obtain
\begin{equation}
{\stackrel{\{\,\}}{D}}\left(\Sigma_\alpha - \tau^\gamma{}_\beta e_\alpha
\rfloor K_\gamma{}^\beta\right) \cong \left(e_\alpha\rfloor{\stackrel
{\{\,\}}{R}}_\gamma{}^\beta - {\stackrel{\{\,\}}{\hbox{\L}}}_\alpha
K_\gamma{}^\beta\right)\wedge\tau^\gamma{}_\beta ,\label{conmomC1}
\end{equation}
whereas replacement of ${\hbox{\L}}_\xi$ by the ordinary Lie derivative 
(\ref{lie0}) yields 
\begin{equation}
d\left(\Sigma_\alpha + \tau^\gamma{}_\beta\,e_\alpha\rfloor
\Gamma_\gamma{}^\beta\right) \cong (\ell_{e_\alpha}\vartheta^\beta)
\wedge\Sigma_\beta + (\ell_{e_\alpha}\Gamma_\gamma{}^\beta)\wedge 
\tau^\gamma{}_\beta.\label{conmomC2}
\end{equation}

The three forms of the Noether identity (\ref{conmomC}), (\ref{conmomC1})
and (\ref{conmomC2}) are equivalent. In \cite{NH} the metric-affine 
counterpart of (\ref{conmomC1}) was used to demonstrate that structureless
test particles always move along Riemannian geodesics. 

\section{Currents associated with a vector field}\label{invcurr}

In the two previous sections we have analyzed the Noether identities which
follow from the diffeomorphism symmetry generated by any vector field $\xi$.
The corresponding {\it covariant} currents do not depend on the latter. 
However, there exists a class of {\it invariant} conserved currents which 
are associated with a given (though arbitrary) vector field. A notable 
example is the well known Komar construction \cite{Komar1,Komar2}. As
another example, take a symmetric energy-momentum tensor $T_j{}^i$ (which 
is covariantly conserved in diffeomorphism-invariant theories) and a {\it 
Killing} field $\xi = \xi^i\partial_i$ (that generates an isometry of the 
spacetime). Then $j^i := \xi^jT_j{}^i$ is a conserved current, and a conserved
charge is defined as the integral $\int_S j^i\,\partial_i\rfloor\eta$ over 
an $(n-1)$-hypersurface $S$. Moreover, it is possible to construct a 
conserved current $(n-1)$-form for any solution of a diffeomorphism-invariant 
model even when $\xi$ is not a Killing field \cite{Wald}. Such a current and 
the corresponding charge are scalars under general coordinate transformations.

In this section, we derive globally conserved currents associated with 
a vector field by making use of the {\it third} covariant Lie derivative, 
namely, the Yano derivative (\ref{lie3}). Note that it is defined without 
any connection. 

\subsection{Gravitational current}

We start from the identity (\ref{lieV1}) for the gravitational Lagrangian,
replace the ordinary Lie derivative $\ell_\xi$ with the covariant Yano
derivative ${\cal L}_\xi$. Then using the properties (\ref{lieT}) and
(\ref{lieR}), we find 
\begin{eqnarray}
({\cal L}_\xi\vartheta^\alpha)\wedge E_\alpha - ({\cal L}_\xi T^\alpha)
\wedge H_\alpha - ({\cal L}_\xi R_\alpha{}^\beta)\wedge H^\alpha{}_\beta 
- {\cal L}_\xi V &=& \nonumber\\
= {\cal L}_\xi\vartheta^\alpha\wedge {\cal E}_\alpha + {\cal L}_\xi
\Gamma_\alpha{}^\beta\wedge {\cal C}^\alpha{}_\beta
-\,d{\cal J}^{\rm grav}[\xi] &=& 0, \label{YlieV}
\end{eqnarray}
where we introduced the scalar $(n-1)$-form
\begin{equation}\label{Jgrav1}
{\cal J}^{\rm grav}[\xi] := \xi\rfloor V + {\cal L}_\xi\vartheta^\alpha
\wedge H_\alpha + {\cal L}_\xi\Gamma_\alpha{}^\beta\wedge H^\alpha{}_\beta.
\end{equation}
By making use of the definitions of the Yano derivative (\ref{yframe}) and 
(\ref{yconn}), and taking into account the Noether identity (\ref{n02}), we 
recast this current into the equivalent form
\begin{equation}
{\cal J}^{\rm grav}[\xi] = d\left(\xi^\alpha\,H_\alpha + \Xi_\alpha{}^\beta
H^\alpha{}_\beta\right) + \xi^\alpha\,{\cal E}_\alpha 
+ \Xi_\alpha{}^\beta\,{\cal C}^\alpha{}_\beta.\label{Jgrav}
\end{equation}

\subsection{Matter current}

In complete analogy with the previous subsection, starting now from the 
identity (\ref{vard2}) for the {\it material} Lagrangian, and replacing
the ordinary Lie derivative $\ell_\xi$ with the covariant Yano derivative 
${\cal L}_\xi$, we obtain 
\begin{eqnarray}
{\cal L}_{\xi}\vartheta^\alpha\wedge{\frac{\partial L} 
{\partial\vartheta^\alpha}}+ {\cal L}_{\xi} T^\alpha\wedge{\frac{\partial L} 
{\partial T^\alpha}} + {\cal L}_{\xi} R_{\alpha}{}^{\beta}\wedge{\frac
{\partial L} {\partial R_{\alpha}{}^{\beta}}} && \nonumber\\ 
+{\cal L}_\xi\Psi^A\wedge{\frac{\partial L} {\partial\Psi^A}}
+ {\cal L}_{\xi}D\Psi^A\wedge {\frac{\partial L} {\partial D\Psi^A}}
- {\cal L}_\xi L &=&\nonumber\\
= {\cal L}_\xi\vartheta^\alpha\wedge\Sigma_\alpha + {\cal L}_\xi
\Gamma_\alpha{}^\beta\wedge \tau^\alpha{}_\beta + {\cal L}_\xi\Psi^A
\wedge {\frac{\delta L} {\delta\Psi^A}}
-\,d{\cal J}^{\rm mat}[\xi] &=& 0. \label{YlieL}
\end{eqnarray}
Here the $(n-1)$-form of the matter current is defined by 
\begin{equation}\label{Jmat1}
{\cal J}^{\rm mat}[\xi] := \xi\rfloor L - {\cal L}_\xi\vartheta^\alpha
\wedge {\frac {\partial L}{\partial T^\alpha}} - {\cal L}_\xi\Gamma_\alpha
{}^\beta\wedge {\frac {\partial L}{\partial R_\alpha{}^\beta}} -
{\cal L}_\xi\Psi^A\wedge {\frac {\partial L}{\partial D\Psi^A}}.
\end{equation}
Again using the definitions of the Yano derivative (\ref{yframe}) and 
(\ref{yconn}), together with the Noether identity (\ref{momC}), we 
find that
\begin{equation}\label{Jmat}
{\cal J}^{\rm mat}[\xi] = -\,d\left(\xi^\alpha{\frac {\partial L}
{\partial T^\alpha}} + \Xi_\alpha{}^\beta {\frac {\partial L}
{\partial R_\alpha{}^\beta}}\right) + \xi^\alpha\Sigma_\alpha 
+ \Xi_\alpha{}^\beta\tau^\alpha{}_\beta.
\end{equation}

\subsection{Total current}

Finally, for the coupled system of gravitational and matter fields
described by the total Lagrangian $V^{\rm tot} = V + L$, the diffeomorphism 
invariance of $V^{\rm tot}$ gives rise to the total current $(n -1)$-form
\begin{eqnarray}
{\cal J}[\xi]&:=&{\cal J}^{\rm grav}[\xi] + {\cal J}^{\rm mat}[\xi]\nonumber\\
&=& \xi\rfloor (V + L) - {\cal L}_\xi\Psi^A\wedge {\frac {\partial L}
{\partial D\Psi^A}} \nonumber\\ \label{Jd1}
&& + {\cal L}_\xi\vartheta^\alpha\wedge\left(H_\alpha - {\frac {\partial L}
{\partial T^\alpha}}\right) + {\cal L}_\xi\Gamma_\alpha{}^\beta\wedge\left( 
H^\alpha{}_\beta - {\frac {\partial L}{\partial R_\alpha{}^\beta}}\right).
\end{eqnarray}
By combining (\ref{YlieV}) and (\ref{YlieL}), we verify that the exterior
derivative of this current is 
\begin{equation}
d{\cal J}[\xi]= {\cal L}_\xi\vartheta^{\alpha}\wedge ({\cal E}_\alpha +
\Sigma_\alpha) + {\cal L}_\xi\Gamma_\alpha{}^{\beta}\wedge({\cal C}^\alpha
{}_\beta + \tau^\alpha{}_\beta) + {\cal L}_\xi\Psi^A\wedge 
{\frac{\delta L} {\delta\Psi^A}}.\label{dcurr}
\end{equation} 
Finally, from (\ref{Jgrav}) and (\ref{Jmat}), we find 
\begin{eqnarray}
{\cal J}[\xi] &=& d\left[(\xi\rfloor\vartheta^\alpha)\left(H_\alpha 
- {\frac {\partial L}{\partial T^\alpha}}\right) + \Xi_\alpha{}^\beta 
\left(H^\alpha{}_\beta - {\frac {\partial L}{\partial R_\alpha
{}^\beta}}\right)\right]\nonumber\\
&& +\,(\xi\rfloor\vartheta^\alpha)({\cal E}_\alpha + \Sigma_\alpha) 
+ \Xi_\alpha{}^\beta ({\cal C}^\alpha{}_\beta + \tau^\alpha{}_\beta).
\label{Jd}
\end{eqnarray} 
As follows from (\ref{dcurr}), for solutions of the coupled system of 
gravitational plus matter field equations (\ref{matter}), (\ref{einstein}) 
and (\ref{cartan}), the total current $(n-1)$-form (\ref{Jd1}) is conserved: 
$d{\cal J}[\xi] = 0$ {\it for any} $\xi$. Hence, it is possible to define 
a corresponding conserved charge ${\cal Q}[\xi]$ by integrating ${\cal J}
[\xi]$ over an $(n-1)$-dimensional spatial hypersurface $S$. Moreover, as 
we see from the equation (\ref{Jd}), on the solutions of the field equations 
(\ref{matter})-(\ref{cartan}), this current is expressed in terms of a 
superpotential $(n-2)$-form. As a result, the corresponding charge can 
be computed as an integral over the spatial boundary $\partial S$:
\begin{equation}
{\cal Q}[\xi] := \int\limits_S {\cal J}[\xi] = \int\limits_{\partial S}\left[
(\xi\rfloor\vartheta^\alpha)\left(H_\alpha - {\frac {\partial L}{\partial 
T^\alpha}}\right) + \Xi_\alpha{}^\beta\left(H^\alpha{}_\beta - 
{\frac {\partial L}{\partial R_\alpha{}^\beta}}\right)\right].\label{calq}
\end{equation} 
For the usual case of {\it minimally coupled} matter ${\partial L}/{\partial 
T^\alpha}=0$ and ${\partial L}/{\partial R_\alpha{}^\beta}=0$, and this
expression then simplifies considerably. 

As a historic remark, let us mention that similar constructions for the 
{\it matter current} were described in the literature in the framework of 
the Einstein-Cartan theory \cite{Trautman73,Trautman05,Benn}, and also in 
the context of metric-affine gravity \cite{Hecht92}. However, the 
conservation of those currents was derived assuming a (Killing) symmetry 
of the gravitational configuration. We can recover the latter result as 
follows. If we assume minimal coupling of matter to gravity and choose 
$\xi$ as a generalized Killing vector which satisfies ${\cal L}_\xi
\vartheta^\alpha =0$ (cf. eq. (\ref{Killing})) and ${\cal L}_\xi
\Gamma_\alpha{}^\beta =0$, then (\ref{Jmat}) yields the matter current 
${\cal J}^{\rm mat}[\xi] = \xi^\alpha\Sigma_\alpha + \Xi_\alpha{}^\beta
\tau^\alpha{}_\beta$, which is conserved in view of (\ref{YlieL}) on the 
solutions of the matter field equations. For spinless matter, i.e. 
$\tau^\alpha{}_\beta=0$, it is sufficient for $\xi$ to satisfy the 
usual Killing equation, ${\cal L}_\xi\vartheta^\alpha=0$.
 
In contrast to that, it is worthwhile to stress that in our derivations 
we never made any assumptions concerning the vector field $\xi$. The 
conservation of the current ${\cal J}[\xi]$ and the existence of the 
corresponding charge ${\cal Q}[\xi]$ do not depend on whether $\xi$ is 
a usual/generalized Killing vector or not. Note also that we did not 
specify either the dimension of spacetime or the form of the 
gravitational Lagrangian.

\section{Generalized Lie derivatives and covariant currents}\label{othercurr}

In the previous section, we have applied the Lie derivative ${\cal L}_\xi$ 
in the sense of Yano to obtain invariant conserved currents associated with
a vector field. This is one of our main results. However, the existence of 
arbitrary generalized Lie derivatives (see Appendix~\ref{app1}) opens 
additional possibilities. Let us recall that in Sec.~\ref{GLNM}, in particular 
see eq. (\ref{zero}), we discovered that {\it all} generalized Lie derivatives 
(i.e., for any $(1,1)$ field $B_\alpha{}^\beta$) are admissible for the 
analysis of the conservation laws related to the diffeomorphism symmetry. 
The specific examples for the corresponding Noether identities were considered
in Sec.~\ref{diff}.

So, a natural question arises: Can we find other conserved currents for the 
generalized Lie derivatives (for an arbitrary $B_\alpha{}^\beta$)? The answer
is positive. Without repeating the computations, we just formulate the result. 

Given a generalized Lie derivative $L_\xi$ defined by some $B_\alpha{}^\beta$,
the diffeomorphism invariance of the total Lagrangian $V^{\rm tot}$ of the
coupled gravitational and matter fields associates to a vector field $\xi$ 
(that generates a diffeomorphism) a current $(n -1)$-form  
\begin{eqnarray}
J_L[\xi] &:=& \xi\rfloor (V + L) - L_\xi\Psi^A\wedge {\frac {\partial L}
{\partial D\Psi^A}} \nonumber\\ \label{JdL1}
&& + L_\xi\vartheta^\alpha\wedge\left(H_\alpha - {\frac {\partial L}
{\partial T^\alpha}}\right) + L_\xi\Gamma_\alpha{}^\beta\wedge\left( 
H^\alpha{}_\beta - {\frac {\partial L}{\partial R_\alpha{}^\beta}}\right).
\end{eqnarray}
The subscript ${}_L$ reflects the fact that this current is defined with 
the help of the generalized Lie derivative $L_\xi$. Moreover, this current
has two basic properties: (i) it satisfies 
\begin{equation}
d(J_L[\xi]) = L_\xi\vartheta^{\alpha}\wedge ({\cal E}_\alpha +
\Sigma_\alpha) + L_\xi\Gamma_\alpha{}^{\beta}\wedge({\cal 
C}^\alpha{}_\beta + \tau^\alpha{}_\beta) + L_\xi\Psi^A\wedge 
{\frac{\delta L} {\delta\Psi^A}}.\label{dcurrL}
\end{equation} 
(ii) it admits the representation 
\begin{eqnarray}
J_L[\xi] &=& d\left[(\xi\rfloor\vartheta^\alpha)\left(H_\alpha 
- {\frac {\partial L}{\partial T^\alpha}}\right) + \left(\xi\rfloor
\Gamma_\alpha{}^\beta - B_\alpha{}^\beta\right) 
\left(H^\alpha{}_\beta - {\frac {\partial L}{\partial R_\alpha
{}^\beta}}\right)\right]\nonumber\\
&& +\,(\xi\rfloor\vartheta^\alpha)({\cal E}_\alpha + \Sigma_\alpha) 
+ \left(\xi\rfloor\Gamma_\alpha{}^\beta - B_\alpha{}^\beta\right) 
({\cal C}^\alpha{}_\beta + \tau^\alpha{}_\beta).\label{JdL}
\end{eqnarray} 
Consequently, for solutions of the field equations (\ref{matter}), 
(\ref{einstein}) and (\ref{cartan}), the new currents are conserved, 
$d(J_L[\xi]) = 0$, and the corresponding charges can be computed as 
an integral over the spatial boundary:
\begin{equation}
Q_L[\xi] := \int\limits_S J_L[\xi] = \int\limits_{\partial S}\left[
(\xi\rfloor\vartheta^\alpha)\left(H_\alpha - {\frac {\partial L}{\partial 
T^\alpha}}\right) + \left(\xi\rfloor\Gamma_\alpha{}^\beta - B_\alpha
{}^\beta\right)\left(H^\alpha{}_\beta - {\frac {\partial L}{\partial 
R_\alpha{}^\beta}}\right)\right].\label{QL}
\end{equation}

When $B_\alpha{}^\beta = -\Theta_\alpha{}^\beta$, we recover 
(\ref{Jd1})--(\ref{calq}) in view of (\ref{Xia}). Below we describe 
a number of additional particular cases. 

\subsection{Covariant current for ``natural'' covariant Lie derivative}

The covariant Lie derivative ${\hbox{\L}}_\xi = D\xi\rfloor + \xi\rfloor D$
is defined by the dynamical connection $\Gamma_\alpha{}^\beta$ of the 
gravitational theory (hence the name ``natural"). It corresponds to the
choice $B_\alpha{}^\beta = \xi\rfloor\Gamma_\alpha{}^\beta$. As a result, 
the conserved current $J_\Gamma[\xi]$ (which is given by (\ref{JdL1}) with
$L_\xi$ replaced with ${\hbox{\L}}_\xi$) gives rise to the charge
\begin{equation}
Q_\Gamma[\xi] := \int\limits_S J_\Gamma[\xi] = \int\limits_{\partial S}
\left[(\xi\rfloor\vartheta^\alpha)\left(H_\alpha - {\frac {\partial L}
{\partial T^\alpha}}\right)\right].\label{QL1}
\end{equation}
This quantity is nontrivial only when the gravitational Lagrangian $V$ 
depends on the torsion, and/or for nonminimal coupling of matter. For all 
other cases this charge is identically zero. For example, for the usual
Einstein-Cartan theory we have $H_\alpha = 0$, and for the minimal coupling
the derivative ${\partial L}/{\partial T^\alpha} =0$ also vanishes, hence
the above charge is trivial. 

\subsection{Covariant current for Riemannian covariant Lie derivative}

The Riemannian covariant Lie derivative ${\stackrel{\{\,\}}{\hbox{\L}}}_\xi
= {\stackrel{\{\,\}}{D}}\xi\rfloor + \xi\rfloor{\stackrel{\{\,\}}{D}}$ is 
defined by the Christoffel connection ${\stackrel{\{\,\}}{\Gamma}}_\alpha
{}^\beta$ (hence the notation with the subscript ${}_{\{\,\}}$). It 
corresponds to the choice $B_\alpha{}^\beta = \xi\rfloor{\stackrel{\{\,\}}
{\Gamma}}_\alpha{}^\beta$. The corresponding current $J_{\{\,\}}[\xi]$ is 
obtained by substituting $L_\xi$ with ${\stackrel{\{\,\}}{\hbox{\L}}}_\xi$ 
in (\ref{JdL1}). It yields the charge
\begin{equation}
Q_{\{\,\}}[\xi] := \int\limits_S J_{\{\,\}}[\xi] = \int\limits_{\partial S}
\left[(\xi\rfloor\vartheta^\alpha)\left(H_\alpha - {\frac {\partial L}
{\partial T^\alpha}}\right) - \xi\rfloor 
K_\alpha{}^\beta\left(H^\alpha{}_\beta 
- {\frac {\partial L}{\partial R_\alpha{}^\beta}}\right)\right].\label{QL2}
\end{equation} 
Here $K_\alpha{}^\beta$ is the contortion, see (\ref{K}). As a result, this
charge reduces to (\ref{QL1}) for all torsion-free solutions, as is
the case, for example, in the standard general relativity theory. 

\subsection{Covariant current for ``background" covariant Lie derivative}

Yet another option arises when we introduce a nondynamical ``background" 
connection $\overline{\Gamma}_\alpha{}^\beta$ (that is different from both
$\Gamma_\alpha{}^\beta$ and ${\stackrel{\{\,\}}{\Gamma}}_\alpha{}^\beta$) and 
define a generalized Lie derivative with the help of $B_\alpha{}^\beta = \xi
\rfloor\overline{\Gamma}_\alpha{}^\beta$. Then the key combination is $\xi
\rfloor\Gamma_\alpha{}^\beta - B_\alpha{}^\beta = \xi\rfloor\Delta
\Gamma_\alpha{}^\beta$, and the corresponding conserved charge reduces to
\begin{equation}\label{QL3}
Q_\Delta[\xi] := \int\limits_S J_\Delta[\xi] = \int\limits_{\partial S}\left[
(\xi\rfloor\vartheta^\alpha)\left(H_\alpha - {\frac {\partial L}{\partial 
T^\alpha}}\right) + \xi\rfloor\Delta\Gamma_\alpha{}^\beta\left(H^\alpha
{}_\beta - {\frac {\partial L}{\partial R_\alpha{}^\beta}}\right)\right].
\end{equation}
The difference $\Delta\Gamma_\alpha{}^\beta := \Gamma_\alpha{}^\beta - 
\overline{\Gamma}_\alpha{}^\beta$ normally should be chosen so that to 
provide a finite value for integral over the spatial boundary. Similar
constructions for the computations of conserved quantities in the 
gauge theories of gravity were used in
\cite{Nester1,Nester2,CN99,Nester3,Hecht,Aros06,conserved}.

\subsection{Noncovariant current for ordinary Lie derivative}

Finally, one can also consider the case when $B_\alpha{}^\beta =0$. Then
the generalized Lie derivative reduces to the ordinary one, $L_\xi =\ell_\xi$.
As a result, we find the following charge 
\begin{equation}
Q_0[\xi] := \int\limits_S J_0[\xi] = \int\limits_{\partial S}\left[
(\xi\rfloor\vartheta^\alpha)\left(H_\alpha - {\frac {\partial L}{\partial 
T^\alpha}}\right) + \xi\rfloor\Gamma_\alpha{}^\beta\left(H^\alpha{}_\beta -
{\frac {\partial L}{\partial R_\alpha{}^\beta}}\right)\right].\label{QL4}
\end{equation} 
Due to the noncovariant character of $\ell_\xi$, this quantity is {\it not 
invariant} under local Lorentz transformations, i.e. $Q_0[\xi]$ depends in 
general on the choice of frame on the spatial boundary $\partial S$. The 
general expression (\ref{QL4}) reduces to the result found by Aros et al. 
for the specific Lagrangian $V$ considered in \cite{A00a}, see also 
\cite{A00b,Aros06b}.

\section{Einstein(-Cartan) gravity}\label{ECT}

Komar \cite{Komar1} gave a formula for the computation of the 
gravitational energy in Einstein's general relativity theory which
proved to give correct (i.e., physically reasonable) results
for asymptotically flat configurations. Later \cite{Komar2} it was 
recognized that this formula also yields the angular momentum for  
rotating configurations. Yet, it remained unclear (for us, at least) 
how this nice formula fits into a general Noether scheme. In the few 
relevant studies, the explanations were based on an assumption that 
$\xi$ is a Killing vector either of the physical spacetime geometry or 
of a background geometry \cite{Benn,Chrusciel,Katz,Petrov,Fer90,Fer94}. 
One of the motivations for the current work was to clarify the status 
of the Komar construction. 

Let us consider the Einstein-Cartan theory that is described by the 
Hilbert-Einstein Lagrangian plus, in general, a cosmological term:
\begin{equation}\label{HEV}
V = - \,{\frac 1{2\kappa}}\left(R^{\alpha\beta}\wedge\eta_{\alpha\beta}
- 2\lambda\eta\right).
\end{equation} 
Here $\kappa$ is the gravitational coupling constant, and $\lambda$ is the 
cosmological constant (with a dimension of the inverse length square). 
Making use of (\ref{HH}), (\ref{EE}) and (\ref{n02}), we find explicitly
\begin{eqnarray}
H_\alpha &=& 0, \qquad E_\alpha = -\,{\frac {1}{2\kappa}}\left(R^{\beta\gamma}
\wedge\eta_{\alpha\beta\gamma} - 2\lambda\eta_\alpha\right),\label{HEa}\\
H_{\alpha\beta} &=& {\frac{1}{2\kappa}}\,\eta_{\alpha\beta}, \qquad 
E_{\alpha\beta}=0.\label{HEab}
\end{eqnarray}
We assume minimal coupling of matter, so that ${\partial L}/{\partial 
T^\alpha} =0$ and ${\partial L}/{\partial R_\alpha{}^\beta} =0$. The 
covariant current (\ref{Jd1}) then reduces to
\begin{equation}
{\cal J}[\xi] = \xi\rfloor (V + L) - {\cal L}_\xi\Psi^A\wedge {\frac 
{\partial L}{\partial D\Psi^A}} + {\frac{1}{2\kappa}}\,{\cal L}_\xi
\Gamma^{\alpha\beta}\wedge\eta_{\alpha\beta}.
\end{equation}
This current is conserved, $d{\cal J}[\xi] =0$, on the solutions of the field 
equations (\ref{matter}), (\ref{einstein}) and (\ref{cartan}), which now read
\begin{eqnarray}
{\frac {1}{2}}\,R^{\beta\gamma}\wedge\eta_{\alpha\beta\gamma} 
- \lambda\eta_\alpha &=& \kappa\Sigma_\alpha,\label{einstein1}\\
{\frac {1}{2}}\,T^{\gamma}\wedge\eta_{\alpha\beta\gamma} &=& 
\kappa\tau_{\alpha\beta},\label{cartan1}\\
{\frac {\delta L}{\delta \Psi^A}} &=& 0.\label{matter1}
\end{eqnarray}
On the other hand, ``on-shell" from (\ref{Jd}) we read off 
\begin{equation}
{\cal J}[\xi] = {\frac{1}{2\kappa}}d\left\{{}^\ast\left[dk+\xi\rfloor
(\vartheta^\lambda\wedge T_\lambda)\right]\right\}, 
\end{equation} 
where we used (\ref{HEa}),
(\ref{HEab}) 
and (\ref{Xi}) (recall that $k:=\xi_\alpha\,\vartheta^\alpha$ is the 1-form 
dual to the vector $\xi$). For the solutions of (\ref{cartan1}) for 
$\tau^\alpha{}_\beta = 0$, i.e. for spinless matter or in vacuum, the 
torsion vanishes, $T^\alpha =0$, and hence the total charge (\ref{calq}) 
finally reduces to 
\begin{equation}
{\cal Q}[\xi] =\frac{1}{2\kappa}\int\limits_{\partial S}
{}^\ast dk.\label{truekomar}
\end{equation}
This invariant conserved quantity ${\cal Q}[\xi]$ is precisely the 
($n$-dimensional generalization of the) Komar formula.

\section{Relocalization of the currents}\label{relocalization}

All our constructions are invariant under coordinate and local Lorentz
transformations. However, besides the local coordinate and the local
Lorentz freedom, there is another ambiguity in the definition of the
conserved quantities. Namely, the field equations always allow for
a {\it relocalization} of the gravitational field momenta. As a result,
the conserved currents and the values of the total charges can be 
changed by means of the relocalization of a translational and
rotational momenta. 
 
More specifically, we consider here the case when a relocalization is
produced by the change of the gravitational field Lagrangian by a total
derivative:
\begin{equation}
V' = V + d\Phi, \qquad \Phi=\Phi(\vartheta^\alpha,\Gamma_\alpha{}^\beta,
T^\alpha,R_\alpha{}^\beta).
\end{equation} 
The term $d\Phi$ changes only the boundary part of the action, leaving
the field equations unchanged. We will assume a boundary $(n-1)$-form 
$\Phi$ whose general variation can be written as
\begin{equation}
 \delta\Phi=\delta\vartheta^\alpha\wedge\frac{\partial\Phi}{
\partial\vartheta^\alpha} 
+\delta\Gamma_\alpha{}^\beta\wedge\frac{\partial\Phi}{
\partial\Gamma_\alpha{}^\beta}+\delta T^\alpha\wedge\frac{\partial\Phi}{
\partial T^\alpha}+\delta R_\alpha{}^\beta\wedge\frac{\partial\Phi}{
\partial R_\alpha{}^\beta}. \label{deltaphi}
\end{equation} 
Then, taking the exterior derivative of (\ref{deltaphi}) and expressing
the variations of $d\vartheta^\alpha$, $dT^\alpha$, $d\Gamma_\alpha{}^\beta$
and $dR_\alpha{}^\beta$ in terms of the variations of
$\vartheta^\alpha$, $T^\alpha$, $\Gamma_\alpha{}^\beta$
and $R_\alpha{}^\beta$, we find
\begin{eqnarray}
{\frac{\partial d\Phi}{\partial\vartheta^\alpha}} &=& - d{\frac 
{\partial\Phi} {\partial\vartheta^\alpha}} + \Gamma_\alpha{}^\beta
\wedge{\frac{\partial\Phi} {\partial\vartheta^\beta}} + R_\alpha{}^\beta
\wedge{\frac{\partial\Phi} {\partial T^\beta}},\label{d1}\\
{\frac{\partial d\Phi}{\partial T^\alpha}} &=& d{\frac{\partial\Phi}
{\partial T^\alpha}} - \Gamma_\alpha{}^\beta\wedge{\frac{\partial\Phi}
{\partial T^\beta}} + {\frac{\partial\Phi}{\partial \vartheta^\alpha}},
\label{d2}\\
{\frac{\partial d\Phi}{\partial R_\alpha{}^\beta}} &=& d{\frac
{\partial\Phi}{\partial R_\alpha{}^\beta}} + \Gamma_\lambda{}^\alpha
\wedge{\frac{\partial\Phi} {\partial R_\lambda{}^\beta}} - \Gamma_\beta
{}^\lambda\wedge{\frac{\partial\Phi} {\partial R_\alpha{}^\lambda}} + 
{\frac{\partial\Phi}{\partial\Gamma_\alpha{}^\beta}} + \vartheta^{
[\alpha}\wedge{\frac{\partial\Phi}{\partial T^{\beta]}}}. \label{d3}
\end{eqnarray} 
We assume that, just like the original Lagrangian $V$, the boundary term 
$d\Phi$ is invariant under Lorentz transformations. Then (making use the 
Lagrange-Noether machinery outlined above) one can verify that
\begin{eqnarray}
{\frac{\partial d\Phi}{\partial\Gamma_\alpha{}^\beta}} &=& - d{\frac{\partial
\Phi} {\partial\Gamma_\alpha{}^\beta}} - \Gamma_\lambda{}^\alpha\wedge
{\frac{\partial\Phi}{\partial\Gamma_\lambda{}^\beta}} + \Gamma_\beta{}^\lambda
\wedge{\frac{\partial\Phi}{\partial\Gamma_\alpha{}^\lambda}} - 
\vartheta^{[\alpha}\wedge{\frac{\partial\Phi}{\partial\vartheta^{\beta]}}}
\nonumber\\
&& - \,T^{[\alpha}\wedge{\frac{\partial\Phi}{ \partial T^{\beta]}}} -R_\lambda
{}^\alpha\wedge{\frac{\partial\Phi}{\partial R_\lambda{}^\beta}} + R_\beta
{}^\lambda\wedge{\frac{\partial\Phi}{\partial R_\alpha{}^\lambda}} \equiv 0.
\end{eqnarray}
Note that unlike the $n$-form $d\Phi$, the $(n-1)$-form $\Phi$ itself does 
not necessarily need to be a scalar under local Lorentz transformations. 
Using (\ref{d1})--(\ref{d3}) in the general definitions (\ref{HH}) and 
(\ref{EE}), we then find the relocalized momenta and gravitational currents:
\begin{eqnarray}
H'_\alpha &=& H_\alpha - d{\frac{\partial\Phi}{\partial T^\alpha}} + 
\Gamma_\alpha{}^\beta\wedge{\frac{\partial\Phi}{\partial T^\beta}} -
{\frac{\partial\Phi}{\partial \vartheta^\beta}},\label{HaR}\\
H'^\alpha{}_\beta &=& H^\alpha{}_\beta - d{\frac{\partial\Phi}{\partial
R_\alpha{}^\beta}} - \Gamma_\lambda{}^\alpha\wedge{\frac{\partial\Phi}
{\partial R_\lambda{}^\beta}} + \Gamma_\beta{}^\lambda\wedge{\frac{\partial
\Phi}{\partial R_\alpha{}^\lambda}} - {\frac{\partial\Phi}{\partial
\Gamma_\alpha{}^\beta}€} - \vartheta^{[\alpha}\wedge{\frac{\partial\Phi}
{\partial T^{\beta]}}},\label{HabR}\\
E'_\alpha &=& E_\alpha - d{\frac {\partial\Phi}{\partial\vartheta^\alpha}}
+ \Gamma_\alpha{}^\beta\wedge{\frac{\partial\Phi}{\partial\vartheta^\beta}}
+ R_\alpha{}^\beta\wedge{\frac{\partial\Phi}{\partial T^\beta}},\\ 
\label{EabR} E'_{\alpha\beta} &=& E_{\alpha\beta} + \vartheta_{[\alpha}
\wedge d{\frac{\partial\Phi}{ \partial T^{\beta]}}} - \vartheta_{[\alpha}
\wedge\Gamma_{\beta]}{}^\lambda\wedge{\frac{\partial\Phi}{\partial T^\lambda}}
-\vartheta_{[\alpha}\wedge{\frac{\partial\Phi}{\partial \vartheta^{\beta]}}}.
\end{eqnarray}
Accordingly, the relocalized translational and rotational momenta (\ref{HaR}) 
and (\ref{HabR}) determine the relocalized conserved currents and charges when 
they are substituted into the corresponding formulas (\ref{Jd1})-(\ref{calq}). 

The choice of a boundary term is fairly arbitrary. For example, a ``universal"
relocalization (in the sense that it is available in all spacetime dimensions)
is defined with the help of a ``background'' connection $\overline{\Gamma}
{}_\alpha{}^\beta$. The latter can be introduced as the limit of the dynamical
connection $\Gamma_\alpha{}^\beta$ at spatial infinity $\partial S$, for 
instance, or fixed from other arguments. Given the background connection, 
one can always (for any $n$) add to a Lagrangian $V$ the covariant boundary 
term $\alpha_0d\Phi_0$ with a constant $\alpha_0$ and the $(n-1)$-form
\begin{equation}
\Phi_0:=\eta_{\alpha\beta}\wedge\Delta\Gamma^{\alpha\beta},\qquad
\Delta\Gamma^{\alpha\beta}:=\Gamma^{\alpha\beta}-\overline{\Gamma}
{}^{\alpha\beta}.
\end{equation}
Then the field momenta are relocalized as 
\begin{equation}
H'_\alpha=H_\alpha-\alpha_0\,\eta_{\alpha\beta\lambda}\wedge\Delta
\Gamma^{\beta\lambda}, \qquad H'_{\alpha\beta}=H_{\alpha\beta}
-(-1)^n\,\alpha_0\,\eta_{\alpha\beta}.\label{rel4}
\end{equation}

Alternatively, there is a possibility to make the arbitrariness in the choice
of the boundary term more narrow by restricting one's attention to the 
topological invariants which always can be added to the action. This option,
however, depends on the spacetime dimension. For instance, in four dimensions 
($n = 4$) we can consider a 3-parameter family of boundary forms 
\begin{eqnarray}
\Phi &:=& \alpha_1\Phi_1 + \alpha_2\Phi_2 + \alpha_3\Phi_3, \\
\Phi_1 &:=& T^\alpha\wedge\vartheta_\alpha,\label{CS1}\\
\Phi_2 &:=& \Gamma_\alpha{}^\beta\wedge \left(R_\beta{}^\alpha
+ \frac{1}{3}\,\Gamma_\beta{}^\lambda\wedge\Gamma_\lambda{}^\alpha
\right),\label{CS2} \\
\Phi_3 &:=& \eta_{\alpha\beta\mu\nu}\,\Gamma^{\alpha\beta}\wedge\left(
R^{\mu\nu}+\frac{1}{3}\,\Gamma^{\mu\lambda}\wedge\Gamma_\lambda{}^\nu
\right).\label{CS3}
\end{eqnarray}
These 3-forms correspond to the Nieh-Yan \cite{nieh1,nieh2}, the Pontryagin 
and the Euler topological invariants, respectively. They represent the 
so-called gravitational (translational and rotational) Chern-Simons 3-forms,
see \cite{CS} for more details. Substituting this into 
(\ref{HaR})-(\ref{EabR}) we then obtain a particular (``topological") 
relocalization:
\begin{eqnarray}
H'_\alpha &=& H_\alpha - 2\alpha_1 T_\alpha,\label{Har}\\
H'_{\alpha\beta} &=& H_{\alpha\beta} - \alpha_1\vartheta_\alpha\wedge
\vartheta_\beta+2\alpha_2R_{\alpha\beta} - 2\alpha_3\,\eta_{\alpha\beta
\mu\nu}\,R^{\mu\nu},\label{Habr} \\
E'_\alpha &=& E_\alpha - 2\alpha_1 DT_\alpha ,\label{Ear}\\
E'_{\alpha\beta}&=&E_{\alpha\beta} + 2\alpha_1\vartheta_{[\alpha}\wedge 
T_{\beta]}.\label{Eabr}
\end{eqnarray} 

We will use this specific relocalization in the subsequent discussion of 
the regularization of invariant conserved quantities.

\section{Examples}\label{Ex}

In order to illustrate how our general formalism works, in this section 
we specialize to the case of four-dimensional theories: $n = 4$, 
$\kappa=8\pi G/c^3$. At first, we study the purely Riemannian (without 
torsion) solutions of the Einstein-Cartan theory which have asymptotic 
AdS behavior. After that, we consider similar configurations 
with torsion that arise as solutions in the quadratic Poincar\'e gauge
gravity theory. 

\subsection{Kerr-AdS solution in Einstein-Cartan
theory}\label{secKAdS}

When the cosmological constant $\lambda$ is nontrivial, the Einstein-Cartan
field equations (\ref{einstein1}), (\ref{cartan1}) admit in vacuum the 
generalized Kerr solution with AdS asymptotics ($\lambda<0$). 
We use a spherical local coordinate system $(t,r,\theta,\varphi)$, and 
choose the coframe as 
\begin{eqnarray}
\vartheta^{\hat 0} &=& \sqrt{\frac{\Delta}{\Sigma}}\left[
cdt-a\Omega\sin^2\theta
\,d\varphi\right],\label{cof0} \\
\vartheta^{\hat 1} &=& \sqrt{\frac{\Sigma}{\Delta}}\, dr,\label{cof1} \\
\vartheta^{\hat 2} &=& \sqrt{\frac{\Sigma}{f}}\, d\theta,\label{cof2} \\
\vartheta^{\hat 3} &=& \sqrt{\frac{f}{\Sigma}}\sin\theta\left[ -a\,cdt
+\Omega (r^2+a^2)\,d\varphi\right].\label{cof3}
\end{eqnarray}
Here the functions and constants are defined by
\begin{eqnarray}
\Delta&:=& (r^2 + a^2)(1-\frac{\lambda}{3}\,r^2) - 2mr,\\
\Sigma&:=& r^2 + a^2\cos^2\theta,\\
f &:=& 1+\frac{\lambda}{3}\,a^2\cos^2\theta,\\
m &:=& \frac{GM}{c^2}, \qquad \Omega:=\frac{1}{1+\frac{\lambda}{3}a^2},
\end{eqnarray}
and $0<t<\infty$, $0<r<\infty$, $0<\theta<\pi$ and $0<\varphi<2\pi$. 
For this solution the curvature has two nonvanishing irreducible
pieces: $R^{\alpha\beta}={}^{(1)}R^{\alpha\beta}+{}^{(6)}R^{\alpha\beta}$ 
(see Appendix~\ref{app2} for definitions of the irreducible parts
of the curvature). The curvature scalar is $R = - 4\lambda$, and the Weyl 
2-form (\ref{Weyl}) has a very special structure. Namely, it is expressed 
only in terms of a certain scalar 2-form $w = {\frac 12}w_{\alpha\beta}
\vartheta^\alpha\wedge\vartheta^\beta$ as follows:
\begin{equation}\label{WeylKads}
W^{\alpha\beta} = \widetilde{w}^{\alpha\beta}\,{}^*w - w^{\alpha\beta}\,w
- {\frac 13}\,{}^\ast(w\wedge{}^\ast w)\,\vartheta^\alpha\wedge\vartheta^\beta 
- {\frac 13}\,{}^\ast(w\wedge w)\,\eta^{\alpha\beta}.
\end{equation}
Here $\widetilde{w}^{\alpha\beta} := {\frac 12}\eta^{\alpha\beta\mu\nu}\,
w_{\mu\nu}$. The 2-form $w$ is given by a simple formula:
\begin{equation}
w = u\,\vartheta^{\hat 0}\wedge\vartheta^{\hat 1} + v\,\vartheta^{\hat 2}
\wedge\vartheta^{\hat 3},
\end{equation}
where its two nontrivial components $u:= w_{{\hat 0}{\hat 1}}$ and $v:= 
w_{{\hat 2}{\hat 3}}$ are functions of $r$ and $\theta$. Their explicit
form is not of interest since we will need only the two invariants
\begin{eqnarray}
{}^\ast(w\wedge{}^\ast w) &=& {\frac {3mr\left(r^2 - 3a^2\cos^2\theta\right)}
{\Sigma^3}},\label{ww1}\\
{}^\ast(w\wedge w) &=& {\frac {3ma\cos\theta\left(3r^2 - a^2\cos^2\right)} 
{\Sigma^3}}.\label{ww2}
\end{eqnarray}
One can straightforwardly find $u$ and $v$ from these two equations, but as
we said, we do not need these functions explicitly.

For a vector field $\xi = \xi^i\partial_i$ with constant holonomic components, 
$\xi^i$, in the coordinate system used in (\ref{cof0})--(\ref{cof3}), the 
computation of conserved charges is fairly straightforward. Substituting 
(\ref{dk}) into (\ref{truekomar}), and using (\ref{dkA})-(\ref{chi}), we 
find in the asymptotically AdS case ($\lambda < 0$):
\begin{equation}
{\cal Q}[\xi] = \xi^0\left[ \frac{\Omega Mc^2}{2} - \frac{4\pi
\Omega c\lambda}{3\kappa}\,r_\infty(r_\infty^2 + a^2)\right] -
\xi^3\,\Omega^2Mca.\label{div}
\end{equation}
Here $r_\infty$ is the radius of the spatial boundary sphere $\partial S$. 
It is worthwhile to note that ${\cal Q}[\partial_r]= 0$ and 
${\cal Q}[\partial_\theta] = 0$. 

When $\lambda = 0$, we recover the usual Komar result with ${\cal Q}
[\partial_t]= Mc^2/2$ and ${\cal Q}[\partial_\varphi] = - Mca$.
For a nontrivial negative cosmological constant, the conserved charge 
${\cal Q}[\partial_\varphi] = -\Omega^2Mca$ is finite, but 
${\cal Q}[\partial_t]$ diverges as $r_\infty\rightarrow\infty$. 
Hence, a regularization is needed.

\subsection{Regularization via relocalization}

In a recent paper \cite{conserved}, we have demonstrated that total
conserved quantities can be regularized by means of a relocalization
of the gravitational field momenta. In Sec.~\ref{relocalization}, we 
discussed a specific relocalization generated by a boundary term in the
action. Here we will use the same method to remove the divergence of the
conserved charge (\ref{div}) for the Kerr-AdS configuration.

A straightforward inspection shows that one can solve the regularization
problem with the help of the relocalization (\ref{Har})-(\ref{Eabr}) 
generated by the Chern-Simons boundary terms (\ref{CS1})-(\ref{CS3}).
More exactly, we can verify that the translational and the rotational
Chern-Simons forms (\ref{CS1}) and (\ref{CS2}) do not affect the total 
charge, whereas the Euler boundary term (\ref{CS3}) does the job. Hence 
we put $\alpha_1 = \alpha_2 = 0$, and consider the relocalization 
$H_{\alpha\beta}\rightarrow H'_{\alpha\beta} = H_{\alpha\beta} - 2
\alpha_3\eta_{\alpha\beta\mu\nu}R^{\mu\nu}$ that is generated by the 
change of the Lagrangian $V\rightarrow V' = V + \alpha_3d\Phi_3$ by the 
boundary term (\ref{CS3}). Using (\ref{HEab}) and the irreducible 
decomposition of the curvature (\ref{curv1}), we find 
\begin{equation}
H'_{\alpha\beta} = \left({\frac 1 {2\kappa}} - {\frac {4\alpha_3
\lambda}3}\right)\eta_{\alpha\beta} - 2\alpha_3\eta_{\alpha\beta\mu\nu}
\overline{R}^{\mu\nu},\label{relH}
\end{equation}
with $\overline{R}^{\alpha\beta} := R^{\alpha\beta} - {\frac \lambda 3}
\,\vartheta^\alpha\wedge\vartheta^\beta$. The term $\eta_{\alpha\beta}$ 
contributes to ${\cal Q}'[\xi]$ with the usual Komar expression ${}^\ast 
dk$ which makes the conserved charge (\ref{div}) infinite. Hence we choose
$\alpha_3 = 3/8\kappa\lambda$ and eliminate the corresponding term completely.
As a result, we end with the regularized invariant conserved charge
\begin{equation}
{\cal Q}'[\xi] = \int\limits_{\partial S}\Xi^{\alpha\beta}H'_{\alpha\beta}
= -\,{\frac 3 {4\kappa\lambda}}\int\limits_{\partial 
S}\eta_{\alpha\beta\mu\nu}
(e^\alpha\rfloor D\xi^\beta)\,\overline{R}^{\mu\nu}\,.\label{genregQ}
\end{equation}
On the Kerr-AdS solution, the relocalized momentum (\ref{relH}) is constructed 
in terms of the Weyl 2-form
\begin{equation}
H'_{\alpha\beta} = -\,{\frac 3 {4\kappa\lambda}}\,\eta_{\alpha\beta\mu\nu}
\,W^{\mu\nu}.\label{Habprime}
\end{equation}
Making use of (\ref{Xi}), (\ref{dk}), and of the explicit form of the 
Weyl 2-form for the Kerr-AdS spacetime, c.f. eq. (\ref{WeylKads}), we derive
\begin{equation}\label{WeylXi}
\eta_{\alpha\beta\mu\nu}\Xi^{\alpha\beta}W^{\mu\nu} = {\frac 23}\left[
{}^\ast(w\wedge{}^\ast w)\,{}^\ast(2\omega - \chi) - {}^\ast(w\wedge w)
\,(2\omega - \chi)\right].
\end{equation}
Substituting now (\ref{ww1}), (\ref{ww2}) and (\ref{dk})-(\ref{chi}), we
finally calculate the regularized charge:
\begin{equation}
{\cal Q}'[\xi] = \xi^0\,\Omega Mc^2 - \xi^3\,\Omega^2Mca\,.\label{regQ}
\end{equation}
In other words, we found finite values of the covariant total charge 
$\cal Q$ for the Kerr-AdS solution, which  read 
\begin{equation}\label{QQ}
{\cal Q}'[\partial_t] = \Omega Mc^2, \quad {\cal Q}'[\partial_\varphi] =
-\Omega^2Mca,
\quad {\cal Q}'[\partial_r] = 0,\quad {\cal Q}'[\partial_\theta] = 0.
\end{equation}	
Our results agree with those in \cite{Hecht93,HT85}.
It is worthwhile to note that the relocalized gravitational Lagrangian 
for $\alpha_3 = 3/8\kappa\lambda$ can be written as
\begin{equation}
V' = {\frac {3}{8\kappa\lambda}}\,\eta_{\alpha\beta\mu\nu}
\overline{R}^{\alpha\beta}\wedge \overline{R}^{\mu\nu}.
\end{equation}
This Lagrangian was studied extensively in the framework of various 
approaches to gravity on the basis of the de Sitter group, see 
\cite{Macdowell,Zardecki,Gotzes,app}, for example. The same action
was also used in \cite{A00a} and \cite{A00b} for the derivation of the
conserved current associated with a vector field. 

Our results (\ref{QQ}) agree with those in \cite{A00a,A00b}
for this particular case (note that the time-like Killing vector used 
in \cite{A00a} corresponds to $\Omega\partial_\varphi$ in our notation). 
However, such an agreement appears to be a mere (although remarkable) 
coincidence since in \cite{A00a} the authors consider the {\it noncovariant}
 charge (\ref{QL4}), which depends in general on the  choice of a frame 
 at spatial infinity.

In order to illustrate this difference, let us take the Kerr-AdS coframe 
(\ref{cof0})-(\ref{cof3}) and evaluate the noninvariant charge $Q_0
[\partial_t]$ 
for a tetrad $\vartheta'^\alpha$ that is obtained from the original one
by means of the local Lorentz transformation  
\begin{eqnarray}
\vartheta'^{\hat 0} &=& \vartheta^{\hat 0}\cosh\zeta(x^i) + \vartheta^{\hat 1}
\sinh\zeta(x^i),\\ 
\vartheta'^{\hat 1} &=& \vartheta^{\hat 0}\sinh\zeta(x^i) + \vartheta^{\hat 1}
\cosh\zeta(x^i),\\
\vartheta'^{\hat 2} &=& \vartheta^{\hat 2},\quad \vartheta'^{\hat 3} 
= \vartheta^{\hat 3}.
\end{eqnarray}
Choosing the function as $\zeta(x^i) = \alpha_0\,rt\sin\theta$ ($\alpha_0$ is 
a constant) we find that eq. (\ref{QL4}) yields $Q'_0[\partial_t] = \Omega 
Mc^2- \alpha_06\pi^2\Omega m/\kappa\lambda$. The value of the charge can thus 
be arbitrary, depending on the constant $\alpha_0$. This charge can even can 
be made divergent. For $\zeta = \alpha_0\, r^2t\sin\theta$, for example, we 
find $Q'_0[\partial_t] = \Omega Mc^2 - r_\infty\alpha_06\pi^2\Omega m/\kappa
\lambda$ which diverges when the spatial boundary is $r_\infty\rightarrow
\infty$. It is certainly true that the parameters of the Lorentz 
transformation
above look quite exotic, so to say. However, this demonstrates, as a matter 
of principle, the fact that $Q_0$ depends on the choice of a frame.

In contrast, our invariant formula (\ref{calq}) yields the same finite value 
for all coordinates and all frames, which is much more advantageous.

\subsection{Kerr-AdS solution with torsion}\label{withT}

In vacuum, the Einstein-Cartan theory coincides with Einstein's general 
relativity theory, and all the solutions are characterized by a vanishing 
torsion. In order to test our approach for the configurations with torsion, 
we will consider a different model, namely, the quadratic Poincar\'e gauge 
theory with the Lagrangian \cite{Heyde1,Heyde2,Rumpf}
\begin{equation}
V = -\,{\frac 1
{2\kappa}}\left[T^\alpha\wedge\vartheta^\beta\wedge{}^\ast\left(
T_\beta\wedge\vartheta_\alpha\right) + {\frac {3}
{4\lambda}}\,R^{\alpha\beta}\wedge
{}^\ast R_{\alpha\beta}\right].\label{Vheyde}
\end{equation} 
This model was extensively studied \cite{Heyde1,Heyde2,Rumpf,Baekler,Hecht93} 
and it was demonstrated that it is a natural generalization of the 
Einstein-Cartan theory. In particular, it was shown that this model has a 
correct Einsteinian limit. Note that we use a slightly different notation 
for the coupling constants in (\ref{Vheyde}), as compared to 
\cite{Heyde1,Heyde2,Rumpf}. Here $\kappa=8\pi G/c^3$ and $\lambda$ has 
dimensions of $(length)^{-2}$.

The translational and the rotational gauge field momenta (\ref{HH}) now read:
\begin{eqnarray} 
H_{\alpha} &=& {\frac {1}{\kappa}}\,\vartheta^\beta\wedge{}^\ast\left(
T_\beta\wedge\vartheta_\alpha\right),\label{HeHa}\\
H_{\alpha\beta} &=& {\frac{3}{4\kappa\lambda}}\,{}^\ast R_{\alpha\beta}.
\label{HeHab}
\end{eqnarray}  
Substituting these expressions into the vacuum field equations ${\cal 
E}_\alpha =0$ and ${\cal C}^\alpha{}_\beta =0$, one can verify that there is 
a generalized Kerr-AdS solution that is described as follows. The coframe is 
again given by the above formulas (\ref{cof0})-(\ref{cof3}), whereas the 
components of the torsion 2-form read:
\begin{eqnarray}
T^{\hat 0} &=& \sqrt{\frac \Sigma \Delta}\left[ -\,v_1\vartheta^{\hat 0}
\wedge\vartheta^{\hat 1} - 2v_4\,\vartheta^{\hat 2}\wedge\vartheta^{\hat 3}
+ \sqrt{\frac \Sigma \Delta}\,{\cal T}\wedge(v_2\vartheta^{\hat 2} + v_3
\vartheta^{\hat 3})\right],\label{T0}\\
T^{\hat 1} &=& T^{\hat 0},\label{T1}\\
T^{\hat 2} &=& \sqrt{\frac \Sigma \Delta}\,{\cal T}\wedge
(v_5\vartheta^{\hat 2} + v_4\vartheta^{\hat 3}),\label{T2}\\
T^{\hat 3} &=& \sqrt{\frac \Sigma \Delta}\,{\cal T}\wedge
( -\,v_4\vartheta^{\hat 2} + v_5\vartheta^{\hat 3}).\label{T3}
\end{eqnarray}
Here we have denoted the functions
\begin{eqnarray}
v_1 &=& {\frac {m(r^2 - a^2\cos^2\theta)}{\Sigma^2}},\quad v_4 = -\,{\frac 
{mra\cos\theta}{\Sigma^2}},\quad v_5 = {\frac {mr^2}{\Sigma^2}},\label{v145}\\
v_2 &=& -\,\sqrt{\frac f\Sigma}\,{\frac {mra^2\sin\theta\cos\theta}
{\Sigma^2}},
\quad v_3 = -\,\sqrt{\frac f\Sigma}\,{\frac {mr^2a\sin\theta}{\Sigma^2}},
\label{v23}
\end{eqnarray}
and introduced the 1-form ${\cal T}:=\vartheta^{\hat 0}-\vartheta^{\hat 1}$.
One can verify that the axial torsion piece vanishes for this configuration, 
while the torsion trace is proportional to the above 1-form, that is:
\begin{equation}
\vartheta_\alpha\wedge T^\alpha = 0,\qquad T = e_\alpha\rfloor T^\alpha = 
-\,{\frac m {\sqrt{\Sigma\Delta}}}\,{\cal T}.\label{Ttrace}
\end{equation}

The Riemann-Cartan curvature 2-form of this solution consists of only two 
irreducible parts (see the definitions in Appendix~\ref{app2}),
\begin{equation}\label{RT01}
R^{\alpha\beta} = {}^{(4)}R^{\alpha\beta} + {}^{(6)}R^{\alpha\beta},
\end{equation}
which read explicitly as follows
\begin{equation}
{}^{(4)}R^{\alpha\beta} = {\frac {\lambda mr}{3\Delta}}
\,{}^{(4)}{\cal R}^{\alpha\beta},\qquad {}^{(6)}R^{\alpha\beta} =
{\frac \lambda 3}\,\vartheta^\alpha\wedge\vartheta^\beta.\label{RT02}
\end{equation}
Here the nonvanishing components of the fourth irreducible part are given by
${}^{(4)}{\cal R}^{{\hat 0}{\hat 2}} = -\,{}^{(4)}{\cal R}^{{\hat 2}{\hat 0}} 
= {}^{(4)}{\cal R}^{{\hat 1}{\hat 2}} = -\,{}^{(4)}{\cal R}^{{\hat 2}{\hat 1}} 
= {\cal T}\wedge\vartheta^{\hat 2}$, and 
${}^{(4)}{\cal R}^{{\hat 0}{\hat 3}} = -\,{}^{(4)}{\cal R}^{{\hat 3}{\hat 0}} 
= {}^{(4)}{\cal R}^{{\hat 1}{\hat 3}} = -\,{}^{(4)}{\cal R}^{{\hat 3}{\hat 1}} 
= {\cal T}\wedge\vartheta^{\hat 3}$.

Substituting all this into (\ref{HeHa}) and (\ref{HeHab}), we explicitly find 
the translational momentum
\begin{eqnarray}
H_{\hat 0} &=& \frac{1}{\kappa}\sqrt{\frac \Sigma \Delta}\left[
-\,2v_4\vartheta^{\hat 0}\wedge
\vartheta^{\hat 1} - 2v_5\,\vartheta^{\hat 2}\wedge\vartheta^{\hat 3}
+ \sqrt{\frac \Sigma \Delta}\,{\cal T}\wedge(v_3\vartheta^{\hat 2} - v_2
\vartheta^{\hat 3})\right],\label{H0}\\
H_{\hat 1} &=& -\,H_{\hat 0},\label{H1}\\
H_{\hat 2} &=& \frac{1}{\kappa}\sqrt{\frac \Sigma \Delta}\,{\cal
T}\wedge\left[-\,v_4
\vartheta^{\hat 2} + (v_1 - v_5)\vartheta^{\hat 3}\right],\label{H2}\\
H_{\hat 3} &=& \frac{1}{\kappa}\sqrt{\frac \Sigma \Delta}\,{\cal
T}\wedge\left[(v_5 - v_1)
\vartheta^{\hat 2} - v_4\vartheta^{\hat 3}\right],\label{H3}
\end{eqnarray}
and the rotational field momentum
\begin{equation}
H^{\alpha\beta} = {\frac 1{4\kappa}}\,\eta^{\alpha\beta} + {\frac {mr}
{4\kappa\Delta}}\,{}^\ast{}^{(4)}{\cal R}^{\alpha\beta}.\label{HTab}
\end{equation}

We consider again vector fields with constant components $\xi^i$ in the 
coordinate system used above. Then, from (\ref{cof0})--(\ref{cof3}) and 
(\ref{H0})--(\ref{H3}), for the translational contribution described by 
the first term in (\ref{calq}) we obtain:
\begin{equation}
\int\limits_{\partial S}(\xi\rfloor\vartheta^\alpha)H_\alpha=\xi^0
\,\left(\Omega Mc^2\right) + \xi^3\left(-\frac{2}{3}\Omega Mac\right).
\label{XiH}
\end{equation} 
On the other hand, taking into account that the axial torsion vanishes
(\ref{Ttrace}), the equation (\ref{Xi}) again yields $\Xi_{\alpha\beta}
= {\frac 12}\,e_\beta\rfloor e_\alpha\rfloor (dk)$. Using then 
(\ref{HTab}), after some algebra we find the second (rotational 
contribution) term in the charge (\ref{calq}):
\begin{eqnarray}
\int\limits_{\partial S}\Xi^{\alpha\beta}H_{\alpha\beta} &=& {\frac{1}
{4\kappa}}\,\int\limits_{\partial S}\Xi^{\alpha\beta}\eta_{\alpha\beta}
+ {\frac{m}{4\kappa}}\,\int\limits_{\partial S}{\frac {r}{\Delta}}
\,\Xi^{\alpha\beta}\,{}^\ast{}^{(4)}{\cal R}_{\alpha\beta} \label{divT0}\\
&=& \left[\xi^0\left(\frac{1}{4}\Omega Mc^2 - {\frac{2\pi\Omega\lambda
c}{3\kappa}}r_\infty(r_\infty^2 + a^2)\right) + \xi^3\left( - \frac{1}{2}
\Omega^2 Mac\right)\right]\nonumber\\ &&\qquad 
+ \,\left[\xi^3\left(- \frac{1}{6}\Omega^2Mac\right)\right].\label{divT}
\end{eqnarray} 
In (\ref{divT}) we have explicitly specified by the square brackets the 
contributions of each integral in (\ref{divT0}). Combining (\ref{XiH}) 
with (\ref{divT}), we obtain the invariant conserved charges 
\begin{equation}\label{div2}
{\cal Q}[\partial_t] = \frac{5}{4}\Omega Mc^2 - {\frac{2\pi\Omega
\lambda c}{3\kappa}}r_\infty(r_\infty^2 + a^2),\qquad 
{\cal Q}[\partial_\varphi]=-\frac{4}{3}\Omega^2Mac,
\end{equation}
whereas again ${\cal Q}[\partial_r] = {\cal Q}[\partial_\theta] =0$. As in 
the Riemannian case discussed in Sec.~\ref{secKAdS}, we thus find a finite 
value for the angular momentum ${\cal Q}[\partial_\varphi]$, but a divergent
energy ${\cal Q}[\partial_t]$.

The source of the divergence is easily detected: From (\ref{divT0}) and 
(\ref{divT}) we can see that it is again the usual Komar term $\Xi^{\alpha
\beta}\eta_{\alpha\beta} = {}^\ast(dk)$ that is responsible for all the
infinite contributions to ${\cal Q}[\partial_t]$. Therefore, we can try to
regularize the conserved charges (\ref{div2}) by relocalizing the field 
momenta in the same way as it was done for the Riemannian case in 
(\ref{relH}). Namely, we add an Euler boundary term, see (\ref{CS3}), to 
the gravitational action (\ref{Vheyde}). In this case the translational 
momentum does not change, $H'_\alpha = H_\alpha$, see (\ref{Har}), but 
the rotational momentum does change, according to (\ref{Habr}). Making use 
of (\ref{RT01}), (\ref{RT02}) and (\ref{HTab}), we then find the 
relocalized Lorentz momentum:
\begin{equation}
H'_{\alpha\beta} = \left({\frac{1}{4\kappa}} - {\frac{4\alpha_3\lambda}{3}}
\right)\eta_{\alpha\beta} + {\frac{mr}{\Delta}}\left({\frac{1}{4\kappa}}
\,{}^\ast{}^{(4)}{\cal R}_{\alpha\beta} - {\frac{2\alpha_3\lambda}{3}}
\,\eta_{\alpha\beta\mu\nu}{}^{(4)}{\cal R}^{\mu\nu}\right).\label{relH2}
\end{equation}
Clearly, the divergent contributions are canceled provided we fix the 
coefficient of the boundary term as $\alpha_3=\frac{3}{16\kappa\lambda}$. 
In addition, we recall the double duality property of the fourth irreducible 
piece of the curvature, namely, ${}^\ast{}^{(4)}R_{\alpha\beta}\equiv 
- {\frac 12}\eta_{\alpha\beta\mu\nu}{}^{(4)}R^{\mu\nu}$, see Eq. (164) of 
\cite{PGrev}, for example. The resulting relocalized quantity then reads
\begin{equation}
H'_{\alpha\beta} = {\frac{mr}{2\kappa\Delta}}\,{}^\ast{}^{(4)}
{\cal R}_{\alpha\beta}.
\end{equation}
Using the explicit expression (\ref{RT02}) for ${}^{(4)}{\cal R}_{\alpha
\beta}$, together with (\ref{Xi}) and (\ref{Ttrace}), we then find
\begin{equation}
\int\limits_{\partial S}\Xi^{\alpha\beta}H'_{\alpha\beta}
= \xi^3\left(- \frac{1}{3}\Omega^2Mac\right) .\label{XiHp}
\end{equation} 
Combining (\ref{XiH}) and (\ref{XiHp}), we finally obtain the regularized
covariant conserved charge:
\begin{equation}
{\cal Q}'[\xi] =
\xi^0\left(\Omega Mc^2\right)+\xi^3\left(-\Omega^2Mac\right).
\end{equation}
It is satisfactory to see that the new current yields the ``standard'' 
values for the energy and the angular momentum, ${\cal
Q}'[\partial_t]=\Omega Mc^2$, 
${\cal Q}'[\partial_\varphi]=-\Omega^2Mac$, as well as the trivial values 
${\cal Q}'[\partial_r]=0$ and ${\cal Q}'[\partial_\theta]=0$. 

It seems worthwhile to mention that although in both examples (without and
with torsion) the resulting values of the regularized charge are the same,
there is an important difference in the way how these values are actually
composed. As compared to the Riemannian case discussed in Sec.~\ref{secKAdS}, 
where the {\it rotational} term $\Xi^{\alpha\beta}H_{\alpha\beta}$ alone gave 
rise both to the total energy and to the total angular momentum, in the 
solution with torsion the whole energy comes only from the {\it translational}
contribution $\xi^\alpha H_\alpha$, whereas the total angular momentum arises
from the combined contributions of the translational and rotational terms.

\section{Discussion and conclusion}\label{Discussion}

In this work, we have analyzed the problem of defining conserved currents 
in diffeomorphism- and local Lorentz-invariant theories. Einstein's 
general relativity theory, Einstein-Cartan theory and general Poincar\'e 
gauge theories of gravity belong to this class of models. We have presented
a systematic derivation of a general expression for conserved currents that
are invariant under both coordinate and local Lorentz transformations. Such
currents ${\cal J}[\xi]$ are associated with a given, but completely 
arbitrary, vector field $\xi$ on the spacetime manifold. The conservation 
law, $d{\cal J}[\xi]=0$, holds ``on-shell", i.e. on every solution of the 
coupled system of the gravitational and matter field equations. Our results
are valid in any spacetime dimension for the field-theoretic models with 
arbitrary Lagrangians. Since the latter are always defined up to a total 
derivative, we discussed the relocalization of the gravitational momenta,
induced by boundary terms in the action, in order to find how such a 
relocalization affects the conserved currents and charges. 

As we stressed, the total invariant current ${\cal J}[\xi]$ is conserved for 
any vector field $\xi$ when the field equations are satisfied. In contrast,
the separate gravitational ${\cal J}^{\rm grav}[\xi]$ and matter ${\cal J
}^{\rm mat}[\xi]$ currents are not conserved in general, not even ``on-shell". 
However, if the vector field $\xi$ generates a symmetry of the field 
configuration, i.e. if the generalized Killing equations are satisfied for
the coframe and connection, ${\cal L}_\xi\vartheta^\alpha=0$ and ${\cal  
L}_\xi\Gamma_\alpha{}^\beta=0$, then (\ref{YlieV}) and (\ref{YlieL}) yield 
two independent conservation laws, $d{\cal J}^{\rm grav}[\xi]=0$ and $d
{\cal J}^{\rm mat}[\xi]=0$. Moreover, in the case of spinless matter 
($\tau^\alpha{}_\beta=0$) or in vacuum, the separate conservation laws 
arise for the case when the vector field $\xi$ is a usual isometry 
(${\cal L}_\xi g_{ij}=0$). In other words, separate invariant conservation
laws exist for the matter and/or gravitational currents under stronger 
(Killing symmetry) conditions on the field configurations, as compared to the 
conservation of the total (gravitational plus matter) current that only 
requires the field equations to be satisfied.

As an immediate consequence of the general results, we have demonstrated
that the Komar construction arises as a particular invariant current 
for the Hilbert-Einstein Lagrangian of the gravitational field. In this
sense, the whole formalism can be viewed as a generalization of the Komar
currents to arbitrary gravitational models with more complicated 
Lagrangians. As we verify, the usual Komar charges diverge for spacetimes 
which are not asymptotically flat (in particular, for the asymptotically 
anti-de Sitter spacetimes). We have shown that the general scheme of 
``regularization via relocalization'' can be used in this case to obtain 
finite total conserved charges. In addition, we considered asymptotically
anti-de Sitter solutions of the quadratic Poincar\'e gauge theory in order 
to test how our general formalism works for models with nontrivial torsion
degrees of freedom. We found divergent total charges, which can again 
be regularized with the help of a relocalization induced by a suitable 
boundary term added to the action. Rather curiously, we found the same 
total final finite charges for the Riemannian and for the quadratic 
Poincar\'e model. This result appears to be quite satisfactory and rather 
nontrivial, since in the quadratic Poincar\'e model the total regularized 
charges arise from different sectors of the dynamical field degrees of 
freedom (translational versus rotational). Moreover, we have verified that
${\cal Q}[\partial_r]={\cal Q}[\partial_\theta]=0$ in both models. 

It seems worthwhile to mention some specific mathematical results obtained
in this study. Namely, we have discovered that the local Lorentz invariance 
of the theory allows for a certain freedom in the description of the 
consequences of the diffeomorphism invariance. This freedom is equivalent 
to a definition of a ``generalized Lie derivative'' that acts on the 
geometrical and matter fields. In other words, there is no unique or 
``natural'' definition of the Lie derivative for the Lorentz-covariant fields
(coframe, torsion and Lorentz-covariant matter fields). For example, it is 
always possible to define the Lie derivative of Lorentz-covariant fields as 
the usual Lie derivative $\ell_\xi$. However, the result will not be a 
Lorentz-covariant field and the conserved currents derived from this 
choice will not be invariant under local Lorentz transformations. We have 
found a variety of consistent covariant Lie derivatives, all leading to 
the definition of invariant conserved currents. Technically,
the latter is a consequence the commuting property of the exterior and
generalized Lie derivative.

There is a number of interesting directions in which we can further develop 
the current formalism. In particular, the general approach can be naturally 
extended to include gravity interacting with gauge fields, thus allowing 
to obtain conserved quantities invariant also under gauge transformations
in addition to the coordinate and local Lorentz invariance. Furthermore, 
it is possible to generalize the current framework to metric-affine 
gravity, so that to include gravitational models with local invariance 
under the general linear group. These developments will be discussed 
elsewhere.

\bigskip
{\bf Acknowledgments}. We would like to thank J.G. Pereira for 
useful discussions and for his hospitality at IFT-UNESP. This work was 
supported by FAPESP (for YNO) and by CNPq (for GFR). We also thank 
F.W. Hehl for the careful reading of the paper and for his numerous
helpful comments.

\appendix

\section{Generalized Lie derivatives}\label{app1}

In this paper, generalized Lie derivatives are heavily used. Here we collect 
the corresponding mathematical definitions and describe properties of the 
generalized Lie derivatives. 

We call an operator $L_\xi$ in the algebra of tensor fields over spacetime
a generalized Lie derivative if it is a {\it derivation} of the tensor algebra
(as defined in \cite{KN63}, e.g.) which satisfies the following properties 
(for any Lorentz-valued $p$-form $\omega^A$, $q$-form $\varphi^A$, and 
constant $\lambda$): 
\begin{enumerate}
\item [(a)] $L_{\lambda\xi}\omega^A=\lambda L_\xi\omega^A $ (linearity),
\item [(b)]
$L_\xi(\omega^A\wedge\varphi^B)=(L_\xi\omega^A)\wedge\varphi^B + \omega^A
\wedge L_\xi\varphi^B$ (Leibniz rule), 
\item [(c)] $L_\xi\phi=\ell_\xi\phi=\xi\rfloor d\phi + d(\xi\rfloor\phi)$ 
(usual Lie derivative for any {\it scalar-valued} $p$-form $\phi$).
\end{enumerate}
In accordance with the general mathematical theory (see Proposition 3.3 
of \cite{KN63}), every derivation is constructed from the ordinary Lie 
derivative $\ell_\xi$ and a $(1,1)$ tensorial field. In this work, we 
explicitly define the generalized Lie derivative by the formula
\begin{equation}
L_\xi\omega^A := \ell_\xi\omega^A + B_\alpha{}^\beta
(\rho^\alpha{}_\beta)^A{}_B\wedge\omega^B\label{genLie}
\end{equation} 
when it acts on any (Lorentz-)covariant $p$-form $\omega^A$. The $(1,1)$ 
field $B_\alpha{}^\beta$ depends linearly on $\xi$ in order to provide 
the above property (a). If, under the local Lorentz transformation 
$\vartheta^\alpha\rightarrow\vartheta'^\alpha = \Lambda^\alpha{}_\beta
\vartheta^\beta$, this field transforms as 
\begin{equation}
 B'_\alpha{}^\beta=(\Lambda^{-1})^\rho{}_\alpha B_\rho{}^\gamma 
\Lambda^\beta{}_\gamma -(\Lambda^{-1})^\gamma{}_\alpha (\xi\rfloor 
d\Lambda^\beta{}_\gamma),\label{TB}
\end{equation} 
then the generalized Lie derivative (\ref{genLie}) of a covariant object 
is again a covariant object with the same transformation properties. In 
addition, we would like to know how $L_\xi$ acts on {\it noncovariant} 
objects, such as $d\omega^A$, for example. In order to find this, it suffices 
to define the Lie derivative of the connection $L_\xi\Gamma_\alpha
{}^\beta$. Then we can compute $L_\xi(d\omega^A)$ by writing $d\omega^A
= D\omega^A-\Gamma_\alpha{}^\beta(\rho^\alpha{}_\beta)^A{}_B\wedge
\omega^B$, and by using the Leibniz rule and the definition (\ref{genLie}).

We define the generalized Lie derivative of the connection by
\begin{equation}
L_\xi\Gamma_\alpha{}^\beta := \ell_\xi\Gamma_\alpha{}^\beta
-\left(dB_\alpha{}^\beta+\Gamma_\lambda{}^\beta B_\alpha{}^\lambda
-\Gamma_\alpha{}^\lambda B_\lambda{}^\beta\right).\label{alt01}
\end{equation} 
If we recall the expression of the curvature 2-form in terms of 
the connection 1-form, we can recast the above definition into an 
equivalent form:
\begin{equation}
L_\xi\Gamma_\alpha{}^\beta = D\left(\xi\rfloor\Gamma_\alpha{}^\beta
-B_\alpha{}^\beta\right) + \xi\rfloor R_\alpha{}^\beta.\label{defLie}
\end{equation} 

Without going into details, let us explain this important point. 
The ordinary Lie derivative commutes with the exterior differential,
$[\ell_\xi,d] = 0$. Using this fact together with the Leibniz rule, 
we can easily compute the Lie derivative of the curvature 2-form:
$\ell_\xi R_\alpha{}^\beta=\ell_\xi\left(d\Gamma_\alpha{}^\beta + 
\Gamma_\lambda{}^\beta\wedge\Gamma_\alpha{}^\lambda\right) = d(
\ell_\xi\Gamma_\alpha{}^\beta) + (\ell_\xi\Gamma_\lambda{}^\beta)
\wedge\Gamma_\alpha{}^\lambda+\Gamma_\lambda{}^\beta\wedge(\ell_\xi
\Gamma_\alpha{}^\lambda) = D(\ell_\xi\Gamma_\alpha{}^\beta)$. Here
we formally write the right-hand side as a ``covariant derivative",
however in reality the resulting expression is {\it not} covariant.  
The generalized Lie derivative (\ref{alt01}) removes this deficiency. 
Indeed, the operation (\ref{genLie}) is well defined on the covariant
curvature 2-form, and the result is the covariant 2-form $L_\xi R_\alpha
{}^\beta$. Now, by applying the covariant differential $D$ to (\ref{alt01}),
or, equivalently, to (\ref{defLie}), we straightforwardly verify that 
$D(L_\xi\Gamma_\alpha{}^\beta) = L_\xi R_\alpha{}^\beta$. In other 
words, in the replacement of the noncovariant $\ell_\xi$ 
with a covariant $L_\xi$, the definition of the generalized Lie 
derivative of the connection (\ref{alt01}) provides the correct
replacement of the formal noncovariant relation $\ell_\xi R_\alpha
{}^\beta = D(\ell_\xi\Gamma_\alpha{}^\beta)$ with an appropriate
covariant formula $L_\xi R_\alpha{}^\beta = D(L_\xi\Gamma_\alpha
{}^\beta)$. 

With these definitions, one can prove that the generalized Lie 
derivative commutes with the exterior derivative, i.e. $\left[L_\xi,
d\right]=0$. For an arbitrary Lorentz-valued form $\omega^A$, we have 
\begin{eqnarray}
L_\xi(d\omega^A)&=&L_\xi\left(D\omega^A-\Gamma_\alpha{}^\beta 
(\rho^\alpha{}_\beta)^A{}_B\wedge\omega^B\right)\nonumber \\
&=& d\ell_\xi\omega^A + \left(\ell_\xi\Gamma_\alpha{}^\beta 
- L_\xi\Gamma_\alpha{}^\beta\right)(\rho^\alpha{}_\beta)^A
{}_B\wedge\omega^B + B_\alpha{}^\beta(\rho^\alpha{}_\beta)^A
{}_B\, d\omega^B \nonumber\\
&& + \,B_\alpha{}^\beta\Gamma_\gamma{}^\delta\left[
(\rho^\alpha{}_\beta)^A{}_C(\rho^\gamma{}_\delta)^C{}_B - 
(\rho^\gamma{}_\delta)^A{}_C(\rho^\alpha{}_\beta)^C{}_B\right]
\wedge\omega^C.
\end{eqnarray} 
Using now (\ref{alt01}) and the commutation relation of the
Lorentz generators,
\begin{equation}
\left[\rho_{\alpha\beta},\rho_{\lambda\rho}\right]^A{}_B = 
{\frac 12}\left[g_{\alpha\rho}\rho_{\lambda\beta}-g_{\lambda\beta}
\rho_{\alpha\rho} + g_{\alpha\lambda}\rho_{\beta\rho} - 
g_{\beta\rho}\rho_{\lambda\alpha}\right]^A{}_B,
\end{equation} 
we ultimately find 
\begin{eqnarray}
L_\xi(d\omega^A) &=& d\ell_\xi\omega^A+dB_\alpha{}^\beta(\rho^\alpha
{}_\beta)^A{}_B\wedge\omega^B+B_\alpha{}^\beta(\rho^\alpha{}_\beta
)^A{}_B\,d\omega^B\nonumber \\
&=& d\ell_\xi\omega^A+d\left(B_\alpha{}^\beta(\rho^\alpha{}_\beta
)^A{}_B\,\omega^B\right) = d(L_\xi\omega^A).
\end{eqnarray} 
Similarly, we can verify that $L_\xi(d\Gamma_\alpha{}^\beta)= d(L_\xi
\Gamma_\alpha{}^\beta)$. The vanishing of the commutator $[L_\xi,d]$ 
on all other geometric quantities follows then directly from these 
formulas and the Leibniz rule.

In conclusion, let us give the following useful identities for the 
basic gravitational field variables:
\begin{eqnarray}
L_\xi \vartheta^\alpha &=&D\xi^\alpha+\xi\rfloor 
T^\alpha+\left(B_\beta{}^\alpha
-\xi\rfloor\Gamma_\beta{}^\alpha\right)\wedge\vartheta^\beta ,\\
L_\xi T^\alpha &=&
D(L_\xi\vartheta^\alpha)+(L_\xi\Gamma_\beta{}
^\alpha)\wedge\vartheta^\beta,\\
L_\xi R_\alpha{}^\beta &=& D(L_\xi\Gamma_\alpha{}^\beta), \\
\left[L_\xi, D\right]\omega^A&=&(L_\xi\Gamma_\alpha{}^\beta)
(\rho^\alpha{}_\beta)^A{}_B\wedge\omega^B.
\end{eqnarray}

\subsection{Yano's derivative}

The covariant Lie derivative in the sense of Yano is defined by 
(\ref{lie3}) and (\ref{theta}). Accordingly, this turns out to
be a particular realization of the generalized Lie derivative that 
corresponds to the choice $B^{\alpha\beta}=- \Theta^{\alpha\beta} 
= -e^{[\alpha}\rfloor\ell_\xi\vartheta^{\beta]}$. 

This choice is special because in a certain sense it is ``minimal''. 
Let us explain this property. The usual Lie derivative $\ell_\xi$ (in 
which case $B_\beta{}^\alpha=0$) of covariant geometrical and matter 
fields is not covariant under local Lorentz transformations. Specifically, 
let us consider the Lie derivative of the coframe. This is a 1-form and 
we can decompose it as follows: $\ell_\xi\vartheta^\alpha=\left(S_\beta
{}^\alpha+A_\beta{}^\alpha\right)\vartheta^\beta$, where $S_\beta
{}^\alpha$ is symmetric, $S_{[\alpha\beta]}=0$, and $A_\beta{}^\alpha$ 
is antisymmetric, $A_{(\alpha\beta)} =0$. Explicitly we find $S_{\alpha
\beta}=e_{(\alpha}\rfloor\ell_\xi\vartheta_{\beta)}\equiv 
h^i_\alpha h^j_\beta\,\ell_\xi g_{ij}/2$ and 
$A_{\alpha\beta}=e_{[\alpha}\rfloor
\ell_\xi\vartheta_{\beta]}$. We can verify that $S_\beta{}^\alpha$ is 
a tensor under local Lorentz transformations, but $A_\beta{}^\alpha$ 
is not. Therefore, the second piece is the source of the noncovariance 
of the usual Lie derivative $\ell_\xi\vartheta^\alpha$ of the coframe.
There exists a unique $B_\beta{}^\alpha$ which introduces a covariant 
Lie derivative $L_\xi\vartheta^\alpha$ 
by just removing the second noncovariant term above, namely 
$B_{\alpha\beta}=-A_{\alpha\beta}=-e_{[\alpha}\rfloor
\ell_\xi\vartheta_ {\beta]}=:-\Theta_{\alpha\beta}$. This definition 
is minimal in the sense that it does not require the choice of any arbitrary
constant. For instance, if $t_\beta{}^\alpha$ is a given \textit{tensor 
field},
then the ``nonminimal'' choice 
$B'_\beta{}^\alpha=-\Theta_\beta{}^\alpha+\alpha\, t_\beta{}^\alpha$ also
leads to a covariant Lie derivative, but introduces an unknown scalar
$\alpha$. In \cite{FFFR99,AFR03,Fer03} the choice
$B_{\alpha\beta}=-\Theta_{\alpha\beta}$ is referred to as ``Kosmann
lift" and it was derived with the help of fairly ad hoc assumptions. 

For this minimal choice we have, for vector-valued forms, connection 
and for matter fields, respectively:
\begin{eqnarray}
{\cal L}_\xi\omega^\alpha &=& \ell_\xi\omega^\alpha -
\Theta_\beta{}^\alpha\,\omega^\beta, \label{yframe}\\
{\cal L}_\xi\Gamma_\alpha{}^\beta &=& D\Xi_\alpha{}^\beta +
\xi\rfloor R_\alpha{}^\beta,\label{yconn}\\
{\cal L}_\xi\Psi^A &=& \ell_\xi\Psi^A - \Theta_\beta{}^\alpha
\,(\rho^\beta{}_\alpha)^A{}_B\Psi^B.\label{ymat}
\end{eqnarray}
Here, in accordance with the general relation (\ref{defLie}), we have
\begin{equation}
\Xi_\alpha{}^\beta :=\xi\rfloor\Gamma_\alpha{}^\beta
+ \Theta_\alpha{}^\beta.\label{Xia}
\end{equation}
For a given vector field $\xi$, we can define a corresponding 1-form
by $k:=(\xi\rfloor\vartheta_\alpha)\vartheta^\alpha = \xi_\alpha
\vartheta^\alpha$. Then a straightforward computation shows that
\begin{eqnarray}
\Xi_{\alpha\beta} &\equiv& e_{[\alpha}\rfloor D\xi_{\beta]}
- \xi\rfloor e_{[\alpha}\rfloor T_{\beta]} \\
&\equiv& {\frac 12}\,e_\beta\rfloor e_\alpha\rfloor\left[
dk + \xi\rfloor (\vartheta^\lambda\wedge T_\lambda)\right].\label{Xi}
\end{eqnarray}

It is worthwhile to calculate explicitly the Yano derivative of the torsion,
curvature, and of the coframe:
\begin{eqnarray}
{\cal L}_\xi T^\alpha &=&
D({\cal L}_\xi\vartheta^\alpha)+({\cal L}_\xi\Gamma_\beta{}
^\alpha)\wedge\vartheta^\beta, \label{lieT}\\
{\cal L}_\xi R_\alpha{}^\beta &=& D({\cal L}_\xi
\Gamma_\alpha{}^\beta), \label{lieR}\\
{\cal L}_\xi \vartheta^\alpha &=&D\xi^\alpha + \xi\rfloor 
T^\alpha-\Xi_\beta{}^\alpha\vartheta^\beta\\ \label{Killing}
&=& \vartheta_\beta\,{\stackrel{\{\,\}}{D}}{}^{(\alpha}\xi^{\beta)}.
\end{eqnarray}
These formulas are quite useful for many practical computations, but with
a word of caution: The two last equations are actually somewhat misleading 
since one might have a wrong impression that a (either non-Riemannian or 
Riemannian) connection is involved. However, the Yano derivative ${\cal 
L}_\xi$ is (like the ordinary Lie derivative $\ell_\xi$) defined independently 
of any connection. This fact becomes more clear if we recall that the 
holonomic form of the second factor in (\ref{Killing}) reads
\begin{equation}
{\stackrel{\{\,\}}{D}}_{(\alpha}\xi_{\beta)} = \frac{1}{2}\,h^i_\alpha 
h^j_\beta\left(\xi^k\partial_k
g_{ij} +  g_{ik}\partial_j\xi^k + g_{kj}\partial_i\xi^k\right).
\end{equation}
As a bonus, from this observation we learn that the Yano derivative of the 
coframe vanishes if and only if $\xi$ is a Killing vector field. 

It is worthwhile to note that in absence of torsion the Yano derivative of the
usual spinor field reduces to the Lie derivative of Kosmann \cite{Kosmann}. 

\subsection{Covariant Lie derivative ${\hbox{\L}}_\xi$}

The covariant Lie derivative (\ref{lie1}) corresponds to the choice 
$B_\alpha{}^\beta=\xi\rfloor\Gamma_\alpha{}^\beta$. Accordingly, the basic 
relations for this case read: 
\begin{eqnarray}
{\hbox{\L}}_\xi \vartheta^\alpha &=&D\xi^\alpha+\xi\rfloor T^\alpha,\\
{\hbox{\L}}_\xi \omega^\alpha&=&\xi\rfloor
D\omega^\alpha+D(\xi\rfloor\omega^\alpha),\\
{\hbox{\L}}_\xi \Psi^A&=&\xi\rfloor
D\Psi^A+D(\xi\rfloor\Psi^A),\\
 {\hbox{\L}}_\xi\Gamma_\alpha{}^\beta&=&\xi\rfloor
R_\alpha{}^\beta.
\end{eqnarray} 

\section{Irreducible decompositions}\label{app2}

In a Riemann-Cartan spacetime, the torsion and the curvature can be 
decomposed into three and six irreducible parts, respectively.

Namely, the torsion 2-form is decomposed as $T^{\alpha}={}^{(1)}T^{\alpha} 
+ {}^{(2)}T^{\alpha} + {}^{(3)}T^{\alpha}$, with
\begin{eqnarray}
{}^{(2)}T^{\alpha}&=& {1\over 3}\vartheta^{\alpha}\wedge (e_\nu\rfloor 
T^\nu),\label{iT2}\\
{}^{(3)}T^{\alpha}&=& -\,{1\over 3}{}^*(\vartheta^{\alpha}\wedge{}^*
(T^{\nu}\wedge\vartheta_{\nu}))= {1\over 3}e^\alpha\rfloor(T^{\nu}\wedge
\vartheta_{\nu}),\label{iT3}\\
{}^{(1)}T^{\alpha}&=& T^{\alpha}-{}^{(2)}T^{\alpha} - {}^{(3)}T^{\alpha}.
\label{iT1}
\end{eqnarray}

The curvature 2-form is decomposed as $R^{\alpha\beta} = 
\sum_{I=1}^6\,{}^{(I)}R^{\alpha\beta}$, with
\begin{eqnarray}
{}^{(2)}R^{\alpha\beta} &=& -\,{}^*(\vartheta^{[\alpha}\wedge
\Psi^{\beta]}),\label{curv2}\\
{}^{(3)}R^{\alpha\beta} &=& -\,{\frac 1{12}}\,{}^*(X
\,\vartheta^\alpha\wedge\vartheta^\beta),\label{curv3}\\
{}^{(4)}R^{\alpha\beta} &=& -\,\vartheta^{[\alpha}\wedge
\Phi^{\beta]},\label{curv4}\\
{}^{(5)}R^{\alpha\beta} &=& -\,{\frac 12}\vartheta^{[\alpha}\wedge 
e^{\beta]}\rfloor(\vartheta^\gamma\wedge R_\gamma),\label{curv5}\\
{}^{(6)}R^{\alpha\beta} &=& -\,{\frac 1{12}}\,R\,\vartheta^\alpha
\wedge\vartheta^\beta,\label{curv6}\\
{}^{(1)}R^{\alpha\beta} &=& R^{\alpha\beta} -  
\sum\limits_{I=2}^6\,{}^{(I)}R^{\alpha\beta}.\label{curv1}
\end{eqnarray}
Here 
\begin{equation}
R^\alpha := e_\beta\rfloor R^{\alpha\beta},\quad R := e_\alpha\rfloor 
R^\alpha,
\quad X^\alpha := {}^*(R^{\beta\alpha}\wedge\vartheta_\beta),\quad 
X := e_\alpha\rfloor X^\alpha,\label{WX}
\end{equation}
and 
\begin{eqnarray}
\Psi_\alpha &:=& X_\alpha - {\frac 14}\,\vartheta_\alpha\,X - {\frac 12}
\,e_\alpha\rfloor (\vartheta^\beta\wedge X_\beta),\label{Psia}\\
\Phi_\alpha &:=& R_\alpha - {\frac 14}\,\vartheta_\alpha\,R - {\frac 12}
\,e_\alpha\rfloor (\vartheta^\beta\wedge R_\beta)\label{Phia}.
\end{eqnarray}

The components of the 2-form $R_\alpha{}^\beta = {\frac 12}R_{\mu\nu\alpha}
{}^\beta\,\vartheta^\mu\wedge\vartheta^\nu$ are identified with the curvature 
tensor $R_{\mu\nu\alpha}{}^\beta$. Accordingly, the Ricci tensor is defined
by the components of the 1-form $R_\alpha = {\rm Ric}_{\beta\alpha}\,
\vartheta^\beta$, where explicitly we have ${\rm Ric}_{\alpha\beta} = 
R_{\gamma\alpha\beta}{}^\gamma$. This tensor is not symmetric, in general.
The curvature scalar is, as usual, $R = g^{\alpha\beta}{\rm Ric}_{\alpha
\beta}$. It determines the 6-th irreducible part (\ref{curv6}) of the 
curvature. {}From (\ref{Phia}) we learn that the 4-th part of the curvature
is given by the symmetric traceless Ricci tensor,   
\begin{equation}
\Phi_\alpha = \left({\rm Ric}_{(\alpha\beta)} - {\frac 14}\,R\,g_{\alpha\beta}
\right)\vartheta^\beta.\label{Ric}
\end{equation}
The first irreducible part (\ref{curv1}) introduces the generalized Weyl 
tensor $C_{\mu\nu\alpha}{}^\beta$ which is defined by the components of 
the Weyl 2-form 
\begin{equation}
W_\alpha{}^\beta := {}^{(1)}R_\alpha{}^\beta = {\frac 12}C_{\mu\nu\alpha}
{}^\beta\,\vartheta^\mu\wedge\vartheta^\nu.\label{Weyl}
\end{equation}
Accordingly, the 1-st, 4-th and 6-th curvature parts reproduce the well-known 
irreducible decomposition of the Riemannian curvature tensor into the Weyl,
traceless Ricci, and curvature scalar parts. The 2-nd, 3-rd and 5-th curvature 
parts are purely non-Riemannian since they all arise from the nontrivial 
right-hand side of the first Bianchi identity $R_\alpha{}^\beta\wedge
\vartheta^\alpha = DT^\beta$, see (\ref{WX}) and (\ref{Psia}).

\section{Vector field in Kerr-AdS spacetime}\label{app3}

Let us take an arbitrary vector field $\xi = \xi^i\partial_i$, the components 
of which are four {\it constant} parameters $\xi^0,\xi^1,\xi^2,\xi^3$. Then 
for the Kerr-AdS spacetime with the coframe given by 
(\ref{cof0})-(\ref{cof3}),
we find for the differential of the corresponding 1-form $k$:
\begin{equation}\label{dk}
dk = \omega + \chi,\quad {\rm with}\quad \omega = 2{\cal A}\vartheta^{\hat 0}
\wedge\vartheta^{\hat 1} + 2{\cal B}\vartheta^{\hat 2}\wedge\vartheta^{\hat 
3}. 
\end{equation}
We have here explicitly the coefficients
\begin{eqnarray}
{\cal A} &=& \xi^0c\left[{\frac {\lambda r}{3}} - {\frac {m(r^2 - a^2\cos^2
\theta)}{\Sigma^2}}\right]\nonumber\\
&& +\,\xi^3a\Omega\sin^2\theta\left[{\frac r\Sigma}\left(1 - {\frac
{\lambda r^2}
{3}}
\right) + {\frac {m(r^2 - a^2\cos^2\theta)}{\Sigma^2}}\right],\label{dkA}\\
{\cal B} &=& \xi^0ca\cos\theta\left[{\frac {\lambda}{3}} + {\frac {2mr}
{\Sigma^2}}\right]\nonumber\\ \label{dkB}
&& -\,\xi^3\Omega\cos\theta\left[{\frac {r^2 + a^2}\Sigma}\left(1 +
{\frac {\lambda 
a^2\cos^2\theta}{3}}\right) + {\frac {2mra^2\sin^2\theta}{\Sigma^2}}\right].
\end{eqnarray}
The 2-form $\chi$ in (\ref{dk}) reads
\begin{eqnarray}
\chi &=& {\frac
{2\sqrt{f\Delta}}{\Sigma}}\Bigg[\xi^3\Omega\sin\theta\left(a\cos
\theta\,\vartheta^{\hat 0}\wedge\vartheta^{\hat 2} + r\vartheta^{\hat 3}
\wedge\vartheta^{\hat 1}\right)\nonumber\\
&& -\,\left(\xi^1{\frac {a^2\sin\theta\cos\theta}\Delta} + \xi^2
{\frac rf}\right)\vartheta^{\hat 1}\wedge\vartheta^{\hat 2}\Bigg].\label{chi}
\end{eqnarray}


\begin{thebibliography}{99}
\bibitem{foot1} These derivatives are defined such 
that $\delta V=\delta\vartheta^\alpha\wedge\frac{\partial V}
{\partial\vartheta^\alpha}+\delta T^\alpha\wedge\frac{\partial V}{\partial 
T^\alpha}+\delta R_\alpha{}^\beta\wedge\frac{\partial V}{\partial 
R_\alpha{}^\beta}$, for arbitrary variations. 
Compare with the conventions in \cite{Kopcz}.
\bibitem{gaugematter}
Our analysis does not include the case of gauge fields (such as 
electromagnetic or Yang-Mills, for example) which are normally described 
as 1-forms on the spacetime manifold. The results of this paper are based 
on the replacement of the ordinary Lie derivatives by generalized
Lie derivatives that are covariant under the local Lorentz group. The
inclusion of gauge matter fields requires a further generalization
to the Lie derivatives covariant under the corresponding gauge group. 
This presents a separate interesting problem which will be considered
elsewhere.
\bibitem{F2} Here and in the following we denote $A^{[\alpha}\wedge
B_{\beta]}:=g^{\alpha\lambda}A_{[\lambda}\wedge
B_{\beta]}$.
\end{thebibliography}
\end{document}